\documentclass[%
 reprint,
superscriptaddress,
 amsmath,amssymb,
 aps,
]{revtex4-2}

\usepackage[utf8]{inputenc}
\usepackage{graphicx}
\usepackage{color}
\usepackage{physics}
\usepackage{braket}
\usepackage{comment}
\usepackage[version=4]{mhchem}
\usepackage[breaklinks=true]{hyperref}
\newcommand{\mr}[1]{\mathrm{#1}}
\date{\today}
\usepackage{amsthm}
\theoremstyle{definition}
\newtheorem{theorem}{Theorem}[]

\newtheorem{proposition}[theorem]{Proposition}
\newtheorem{lemma}[theorem]{Lemma}
\newtheorem*{proposition2*}{Proposition}

\begin{document}
\title{Local variational quantum compilation of a large-scale Hamiltonian dynamics}

\author{Kaoru Mizuta}
\email{mizuta.kaoru.65u@st.kyoto-u.ac.jp}
\affiliation{QunaSys Inc., Aqua Hakusan Building 9F, 1-13-7 Hakusan, Bunkyo, Tokyo 113-0001, Japan}
\affiliation{Department of Physics, Kyoto University, Kyoto 606-8502, Japan}

\author{Yuya O. Nakagawa}
\affiliation{QunaSys Inc., Aqua Hakusan Building 9F, 1-13-7 Hakusan, Bunkyo, Tokyo 113-0001, Japan}

\author{Kosuke Mitarai}
\affiliation{Graduate School of Engineering Science, Osaka University, 1-3 Machikaneyama, Toyonaka, Osaka 560-8531, Japan.}
\affiliation{Center for Quantum Information and Quantum Biology, Osaka University, Japan.}
\affiliation{JST, PRESTO, 4-1-8 Honcho, Kawaguchi, Saitama 332-0012, Japan}

\author{Keisuke Fujii}
\affiliation{Graduate School of Engineering Science, Osaka University, 1-3 Machikaneyama, Toyonaka, Osaka 560-8531, Japan.}
\affiliation{Center for Quantum Information and Quantum Biology, Osaka University, Japan.}
\affiliation{RIKEN Center for Quantum Computing (RQC), Hirosawa 2-1, Wako, Saitama 351-0198, Japan}
\affiliation{Fujitsu Quantum Computing Joint Research Division at QIQB,
Osaka University, 1-2 Machikaneyama, Toyonaka 560-0043, Japan}

\begin{abstract}
Implementing time evolution operators on quantum circuits is important for quantum simulation. However, the standard way, Trotterization, requires a huge numbers of gates to achieve desirable accuracy.
Here, we propose a local variational quantum compilation (LVQC) algorithm, which allows to accurately and efficiently compile a time evolution operators on a large-scale quantum system by the optimization with smaller-size quantum systems.
LVQC utilizes a subsystem cost function, which approximates the fidelity of the whole circuit, defined for each subsystem as large as approximate causal cones brought by the Lieb-Robinson (LR) bound.
We rigorously derive its scaling property with respect to the subsystem size, and show that the optimization conducted on the subsystem size leads to the compilation of whole-system time evolution operators.
As a result, LVQC runs with limited-size quantum computers or classical simulators that can handle such smaller quantum systems.
For instance, finite-ranged and short-ranged interacting $L$-size systems can be compiled with $O(L^0)$- or $O(\log L)$-size quantum systems depending on observables of interest.
Furthermore, since this formalism relies only on the LR bound, it can efficiently construct time evolution operators of various systems in generic dimension involving finite-, short-, and long-ranged interactions.
We also numerically demonstrate the LVQC algorithm for one-dimensional systems. Employing classical simulation by time-evolving block decimation, we succeed in compressing the depth of a time evolution operators up to $40$ qubits by the compilation for $20$ qubits. 
LVQC not only provides classical protocols for designing large-scale quantum circuits, but also will shed light on applications of intermediate-scale quantum devices in implementing algorithms in larger-scale quantum devices.
\end{abstract}

\maketitle

\section{Introduction}\label{Sec:Introduction}

Implementing time evolution operators under a large-scale Hamiltonian is one of the most important tasks in noisy intermediate-scale quantum (NISQ) devices \cite{Preskill2018-uo} and larger fault-tolerant quantum computers to exploit their computational power. 
The task is computationally hard for classical computers; despite the enormous effort toward its efficient computation, it generally takes resources that are exponential to the system size.
On the other hand, quantum computers are capable of executing it in polynomial time \cite{Lloyd1996-ko}.
It is also important for computing eigenvalues and eigenstates of a system on a quantum computer; the quantum phase estimation algorithm \cite{Yu_Kitaev1995-oi,Cleve1998-ou,nielsen2002quantum} uses controlled time evolution operators to generate them.
Recent hardware with tens of qubits has realized its proof-of-principle demonstrations for systems such as Fermi-Hubbard models \cite{Arute2020-qr}, discrete time crystals \cite{Mi2021-dtc,Randall2021-ke}, and various equilibrium and nonequilibrium phenomena \cite{Smith2019-zz,Neill2021-jv,Zhu2021-ws}.

\begin{figure}[t]
    \includegraphics[height=7.25cm, width=8.5cm]{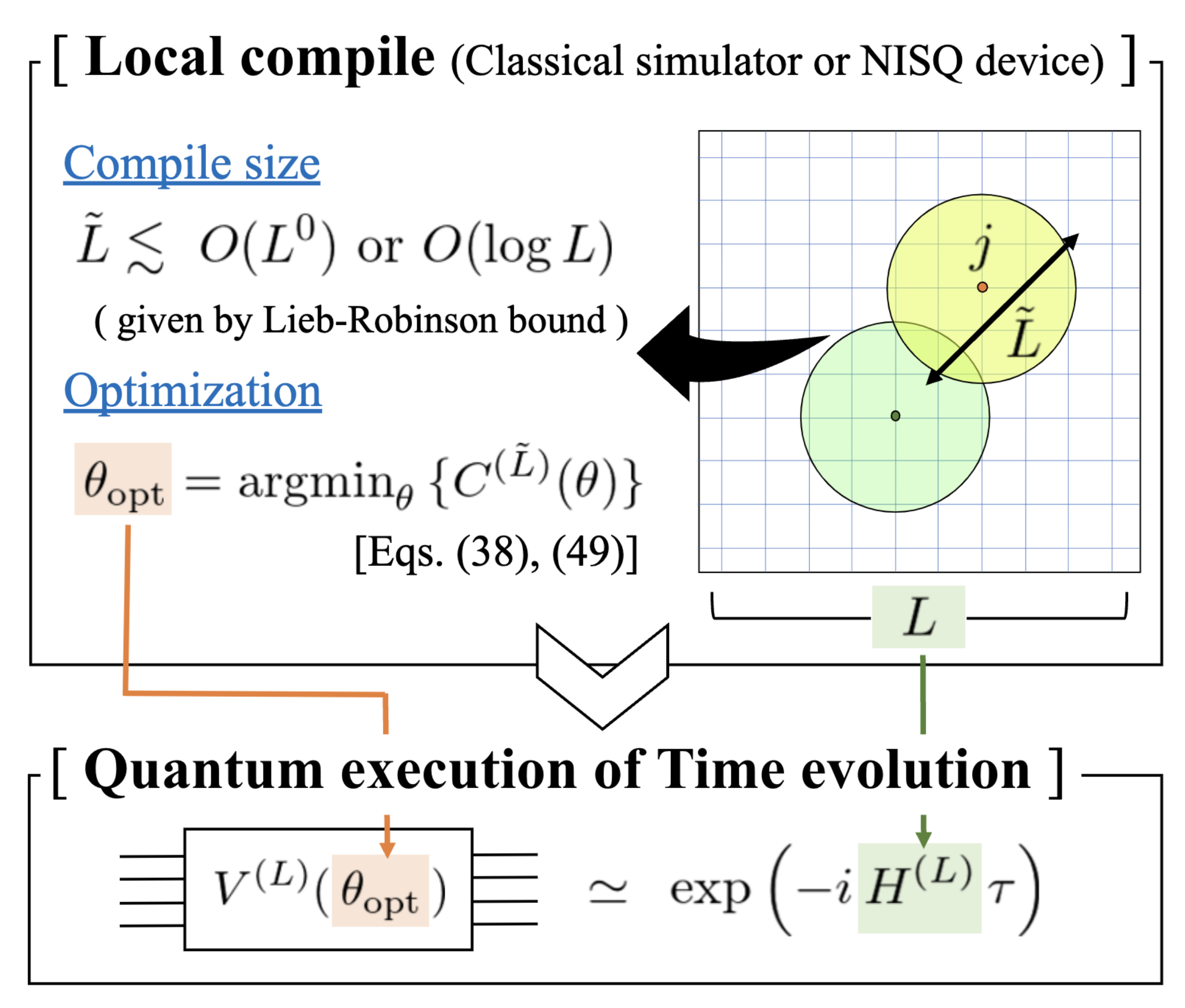}
    \caption{Overview of the local variational quantum compilation (LVQC) protocol. We optimize the cost functions for the compilation size $\tilde{L}$, determined by the Lieb-Robinson (LR) bound. For finite-ranged and short-ranged interacting cases, it typically gives $\tilde{L} \lesssim O(L^0)$ or $\tilde{L} \lesssim O(\log L)$. We can directly implement a large-scale time evolution operator with the optimal parameter $\theta_\mr{opt}$. LVQC can be completed by classical simulation with some approximation or NISQ devices, without implementing the target $\exp (-i H^{(L)} \tau)$ itself.} 
    \label{fig:summary}
\end{figure}

Trotterization is one of the simplest implementations, which has been extensively investigated theoretically \cite{Lloyd1996-ko,Abrams1997-jb,Sornborger1999-oq,Campbell2019-prl,Childs2019fasterquantum,Childs2021-trotter,Ouyang2020compilation} and employed in various experiments such as Refs. \cite{Arute2020-qr,Smith2019-zz,Zhu2021-ws,OMalley2016-dh,Lanyon2011-qo}.
Despite recent developments, it may involve a huge number of gates when applied to a large scale problem with over 50 qubits.
For example, Ref. \cite{Ouyang2020compilation} estimated that we need $10^{15}$-$10^{18}$ gates to perform time evolution of simple molecules.
Even for the simpler Heisenberg model, it is estimated that $10^{6}$-$10^{7}$ elementary rotation gates are needed \cite{Childs2019fasterquantum}.
These estimates are well beyond the reach of current quantum devices whose gate infildelities are on the order of $1$\%.
Moreover, it is problematic even for an ideal fault-tolerant quantum computer because execution of $10^{15}$ gates would require years even if it can perform $10^8$ gates per second. 

It is therefore vital to develop methods that can compress the circuits for time evolution.
The so-called qubitization technique \cite{Low2019-lf} has achieved an optimal scaling in the number of gates needed, but requires many ancilla qubits for its implementation (see e.g. Ref. \cite{Babbush2018-prx}).
When focusing on algorithms that requires no or few ancilla qubits, Refs. \cite{Kokcu2021-on,Gulania2021-eu}, for example, have presented depth-compression methods for Trotter expansion based on some algebraic structures.
Another promising approach is to use the framework of variational quantum algorithms \cite{cerezo2021variational}.
They are exemplified by variational quantum simulation \cite{Li2017-ht,Yuan2019-kq,Heya2019-xp,Endo2020-mp,Benedetti2021-lt,Lin2021-vp,Berthusen2021-qv}, and quantum compilations employing variational quantum diagonalization \cite{Cirstoiu2020-kk,Commeau2020-tf,Gibbs2021-mu}.
Among other methods, quantum-assisted quantum compiling (QAQC) \cite{Khatri2019-gj,Sharma2020-re} and its variant \cite{bilek2022recursive} are one of the promising ways to obtain approximate time evolution operators with compressed circuit depth. 
It uses a variational quantum circuit $V$ to approximate a target unitary $U$. Importantly, they employed a local cost function instead of the naive global fidelity measure $\Tr(U^\dagger V)$ to avoid the barren plateau problem.
While QAQC is available for generic target unitary gate $U$ on $L$ qubits, it seems to be problematic for depth compression that the target $U$ itself should be accurately implemented on quantum circuits.

In this paper, we develop a local variational quantum compilation (LVQC) protocol to search an accurate and efficient quantum circuit for constructing a large-scale local Hamiltonian dynamics with limited-size quantum devices or possibly with classical simulation of such limited-size quantum circuits.
To formulate the protocol, we focus on Lieb-Robinson (LR) bound \cite{Lieb1972-uo}, which dictates that the dynamics under a local Hamiltonian has approximate causal cones.
We compose of subsystem cost functions for every subsystem which measure the local difference between the target unitary gate and the ansatz.
Exploiting the LR bound, we rigorously derive their scaling, which is validated when the subsystem size is as large as the approximate causal cone.
These results lead to our LVQC protocol as described in Fig. \ref{fig:summary}; we optimize a local-compilation cost function, corresponding to the average of the subsystem cost functions over subsystems. 
This cost function can be computed with a at-most $2\tilde{L}$-qubit quantum device or a corresponding classical simulator, where $\tilde{L}$ ($<L$: system size) denotes the scale of the causal cone size. Finally, we construct a quantum circuit that approximates the target time evolution operator for the system size $L$ based on the resulting optimal parameters.

We also conduct classical numerical demonstration of LVQC to compress the depth of the ideal time evolution operators.
We adopt a one-dimensional Heisenberg model, and optimize the cost function for subsystems by approximately computing it with time-evolving block decimation (TEBD) \cite{Vidal2003-eb,Vidal2004-oa}.
We successfully compose of a $5$-depth time evolution operator for $40$ qubits by the local compilation for $20$-qubit systems. This achieves the average gate fidelity $0.9977$, which is much better than that of the same-depth Trotter decomposition, $0.8580$. 
In addition, by computing the stroboscopic dynamics of ferromagnetic states with local excitations or domain walls, the optimal ansatz obtained by LVQC reproduces the dynamics with size- and time-scales twice and ten times as large as those used in the compilation, respectively. 

We emphasize some advantages of LVQC. 
First, it requires at-most $2 \tilde{L}$-qubit quantum devices as large as the causal cone size, which is comparably smaller than the whole-system size $L$. 
There is no need for preparing the ideal target unitary gate for the size $L$ during our protocol.
Second, our formulation relies only on the existence of the LR bounds. LVQC is available for broad systems involving finite-ranged, short-ranged, and long-ranged interactions in generic dimension, with the help of the recent developments in the LR bound \cite{Robinson1976-wd,Nachtergaele2006-ok,Nachtergaele2006-il,Hastings2006-ob,Foss-Feig2015-ca,Matsuta2017-mf,Else2020-tr,Kuwahara2020-se,Tran2021-pv}. 
We expect that LVQC can be applied for executing large-scale time evolution operators in the following ways;
\begin{enumerate}
\item Classical local compilation with approximations \\
\& Quantum execution in NISQ or larger systems
\item Quantum local compilation by NISQ devices \\
\& Quantum execution in larger quantum devices
\end{enumerate}
The first case is exemplified by our numerical results based on TEBD. 
LVQC ensures the small-size compilation sometimes accessible with classical simulators employing some approximations. 
In that case, we can classically compile time evolution operators without suffering noises and statistical errors, and can simulate large-scale quantum systems that are inaccessible only with classical simulators; long-time behaviors beyond the coherence time will be observed in recent programmable quantum simulators by the optimized time evolution operators. 
The second one is rather a long-term perspective. 
To simulate quantum materials with generic dimensions or interactions by NISQ devices or larger fault-tolerant quantum computers, the local-system size for the compilation will become at-least hundreds or thousands of qubits. 
This is just suitable for NISQ devices in the near future, and hence our results will contribute to bridging the gap between NISQ devices and larger-scale quantum computers.

The rest of this paper is organized as follows. 
In Sec. \ref{Sec:Preliminaries}, we introduce QAQC and the LR bound as the preliminaries for our results. 
We devote Sec. \ref{Sec:Scalablity_of_cost}, Sec.  \ref{Sec:Local_compilation} and Sec. \ref{Sec:Numerical} to provide the main results. 
In Sec. \ref{Sec:Scalablity_of_cost}, we introduce the subsystem cost function from the local cost functions of QAQC, and rigorously prove its scaling property by the LR bounds.
In Sec. \ref{Sec:Local_compilation}, we formulate the LVQC protocols respectively for translationally-invariant systems and other generic systems.
The above scaling yields the local compilation of a large-scale Hamiltonian dynamics for both cases, while the protocol is simplified in the former case.
Finally, we show its numerical verification in Sec. \ref{Sec:Numerical} and conclude this paper in Sec. \ref{Sec:Discussion}.

\section{Preliminaries}\label{Sec:Preliminaries}

\begin{figure}[t]
    \includegraphics[height=5.5cm, width=8.5cm]{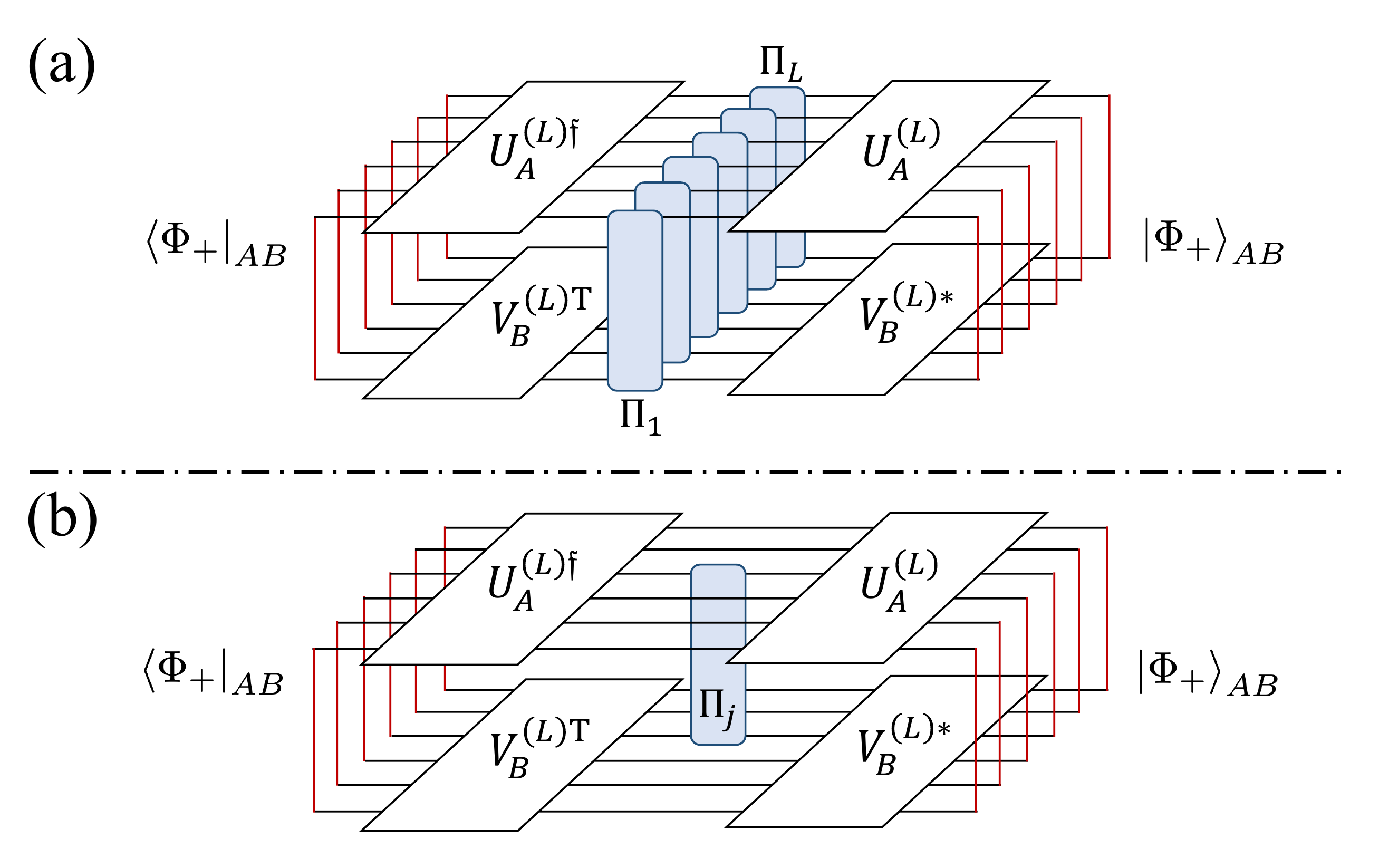}
    \caption{Schematic picture of the way to compute the global and the local cost functions. In each figure, applying a Bell pair $\ket{\Phi_+}_{A_j B_j}$ indicates taking contractions on the $j$-th pair $A_j$ and $B_j$, which we represent by the red solid lines. (a) Schematic picture of $\mr{Tr}[\Pi_1 \hdots \Pi_L \rho_{AB}(U,V)]$, which gives the global cost function $C_\mr{HST}$ via Eq. (\ref{Eq:Cost_HST}). (b) Schematic picture of $\mr{Tr}[\Pi_j \rho_{AB}(U,V)]$, which gives the local cost function $C_\mr{LHST}$ via Eq. (\ref{Eq:Cost_LHST_j}).} 
    \label{fig:preliminary}
\end{figure}

In this section, we review some preliminary studies in order to derive our results on LVQC for a large-scale Hamiltonian dynamics.

\subsection{Quantum-assisted quantum compiling (QAQC)}\label{Subsec:QAQC}
Quantum-assisted quantum compiling (QAQC) \cite{Khatri2019-gj} is a quantum-classical hybrid algorithm to obtain a variational quantum circuit $V(\theta)$ with parameters $\theta$, which approximates a target unitary operator $U$.
They have introduced several cost functions $C(U,V)$, that should be minimized, to to obtain an optimal parameter $\theta_{\mathrm{opt}}$ such that $U \simeq V(\theta_\mr{opt})$.
The cost functions $C(U,V)$ should satisfy the following properties;
\begin{enumerate}
    \item (Computability) We can efficiently compute $C(U,V)$ with a quantum computer.
    \item (Faithfulness) $C(U,V)$ is always non-negative, and it becomes $0$ if and only if $U$ and $V$ are equivalent.
    \item (Operational meaning) $C(U,V)$ provides constraints on some operationally meaningful value.
\end{enumerate}

The first cost function is a global one defined by
\begin{equation}\label{Eq:Cost_HST_Def}
    C_\mr{HST}(U,V) = 1 - \frac{1}{4^L} | \mr{Tr} [U^\dagger V] | ^2,
\end{equation}
when $U$ and $V$ are defined on an $L$-qubit lattice $\Lambda$. This can be measured by Hilbert-Schmidt Test (HST). In HST, we use an $2L$-qubit lattice $\Lambda_A \cup \Lambda_B$ (each of $\Lambda_A$ and $\Lambda_B$ is a copy of $\Lambda$), and initialize the state by the Bell state $\ket{\Phi_+}_{AB}$, defined by
\begin{eqnarray}
     \ket{\Phi_+}_{AB} &=& \bigotimes_{j \in \Lambda} \ket{\Phi_+}_{A_j B_j}, \\
     \ket{\Phi_+}_{A_j B_j} &=& \frac{1}{\sqrt{2}} (\ket{00}+\ket{11})_{A_j B_j}.
\end{eqnarray}
The state $\ket{\Phi_+}_{A_j B_j}$ represents the Bell pair of the $j$-th sites $A_j$ and $B_j$ respectively in $\Lambda_A$ and $\Lambda_B$. Then, we apply $U$ and $V^\ast$ respectively to the subsystems $A$ and $B$, resulting in the state
\begin{equation}
    \rho_{AB}(U,V) = (U_A \otimes V^\ast_B) \ket{\Phi_+}_{AB}\bra{\Phi_+}_{AB} (U_A \otimes V^\ast_B)^\dagger,
\end{equation}
and perform the Bell measurements for every $j$-th pair $A_j$ and $B_j$. It is equivalent to measure $\Pi_1 \Pi_2 \hdots \Pi_L$, where $\Pi_j$ is defined by
\begin{equation}
    \Pi_j = \ket{\Phi_+}_{A_j B_j} \bra{\Phi_+}_{A_j B_j}.
\end{equation}
Finally, since Eq. (\ref{Eq:Cost_HST_Def}) can be rewritten as
\begin{equation}\label{Eq:Cost_HST}
    C_\mr{HST}(U,V) = 1 - \mr{Tr} [\Pi_1 \Pi_2 \hdots \Pi_L \rho_{AB}(U,V)],
\end{equation}
we can efficiently compute $C_\mr{HST}(U,V)$ with a $2L$-qubit quantum device. 
The term $\mr{Tr} [\Pi_1 \Pi_2 \hdots \Pi_L \rho_{AB}(U,V)]$ is schematically depicted by Fig. \ref{fig:preliminary} (a).
The cost function $C_\mr{HST}(U,V)$ is faithful in that it satisfies $0 \leq C_\mr{HST}(U,V) \leq 1$ and that it becomes $0$ if and only if there exists $\varphi \in \mathbb{R}$ such that $U=e^{i \varphi} V$. 

The second one is a local cost function defined by
\begin{equation}\label{Eq:Cost_LHST}
    C_\mr{LHST}(U,V) = \frac{1}{L} \sum_{j=1}^L C_\mr{LHST}^{(j)}(U,V),
\end{equation}
where each term is given by
\begin{equation}\label{Eq:Cost_LHST_j}
    C_\mr{LHST}^{(j)}(U,V) = 1 - \mr{Tr} [\Pi_j \rho_\mr{AB}(U,V)],
\end{equation}
for $j=1,2,\hdots,L$. They satisfy $0 \leq C_\mr{LHST}(U,V) \leq 1$ and $0 \leq C_\mr{LHST}^{(j)}(U,V) \leq 1$ by their definitions. We can compute them on a $2L$-qubit quantum device by Local Hilbert-Schmidt Test (LHST), in which we perform Bell measurement of the $j$-th pair $A_j$ and $B_j$ on the state $\rho_{AB}(U,V)$ for $C_\mr{LHST}^{(j)}(U,V)$ and take its average for $ C_\mr{LHST}(U,V)$.
The term $\mr{Tr} [\Pi_j \rho_\mr{AB}(U,V)]$ is described by Fig. \ref{fig:preliminary} (b).
In terms of faithfulness, $ C_\mr{LHST}^{(j)}(U,V)$ satisfies the following property,
\begin{eqnarray}
    && C_\mr{LHST}^{(j)}(U,V) = 0 \quad \text{if and only if} \nonumber
    \\
    && \quad \text{$\,^\exists \varphi \in \mathbb{R}$, $\,^\exists W$: unitary, s.t. $UV^\dagger = e^{i \varphi} I_{\{j\} } \otimes W$}, \label{Eq:LHST_zero_condition}
\end{eqnarray}
where $I_{\{j\}}$ denotes the identity operator acting on $j$-th qubit.
This indicates that the action of $U$ corresponds to that of $V$ on the $j$-th site. Thus, the cost function $ C_\mr{LHST}(U,V)$ becomes $0$ if and only if there exists $\varphi \in \mathbb{R}$ such that $U=e^{i \varphi} V$.

In QAQC in Ref. \cite{Khatri2019-gj}, the authors employ either or the combined cost function
\begin{equation}
    C_\alpha(U,V) = \alpha C_\mr{HST}(U,V) + (1-\alpha) C_\mr{LHST}(U,V),
\end{equation}
with $0 \leq \alpha \leq 1$. It is faithful, and possesses an operational meaning in terms of the average gate fidelity, defined by
\begin{equation}
    \bar{F}(U,V) = \int_\psi |\braket{\psi | U^\dagger V | \psi}|^2 d \psi, \quad \text{$\psi$: Haar random state}.
\end{equation}
This indicates the expected fidelity between $U \ket{\psi}$ and $V\ket{\psi}$ averaged over a Haar random state $\ket{\psi}$, and it is bounded from below by the resulting cost functions as follows \cite{Horodecki1999-ws,Nielsen2002-ia,Khatri2019-gj},
\begin{eqnarray}
    \bar{F}(U,V) &=& 1 - \frac{2^{|\Lambda|}}{2^{|\Lambda|}+1} C_\mr{HST}(U,V), \label{Eq:operational_meaning_HST} \\
    \bar{F}(U,V) &\geq& 1 - \frac{2^{|\Lambda|}}{2^{|\Lambda|}+1} \cdot |\Lambda| C_\mr{LHST} (U,V), \label{Eq:operational_meaning_LHST}
\end{eqnarray}
where $|\Lambda|$ denotes the number of sites in the lattice $\Lambda$. The cost functions of QAQC can be efficiently computed on a $2L$-qubit quantum device based on Eqs. (\ref{Eq:Cost_HST}) and (\ref{Eq:Cost_LHST_j}).
Alternatively, we can nontrivially reduce the resource for cost evaluation to $L$ qubits by the following lemma, which we prove in Appendix \ref{Asec:LqubitLHST}.
\begin{lemma}\label{Thm:LqubitLHST}
$C_{\mathrm{HST}}(U,V)$ and $C_{\mathrm{LHST}}(U,V)$ for $L$-qubit unitaries $U$ and $V$ can be evaluated effciently within an additive error $\epsilon$ with $\mathcal{O}(1/\epsilon^2)$ runs of an $L$-qubit device.
\end{lemma}
\noindent It should be noted that the algorithm to achieve Lemma \ref{Thm:LqubitLHST} involves a Monte-Carlo sampling and induces increased (however constant) overhead compared to the case where we use $2L$ qubits.
In any cases, the bottleneck of QAQC for compressing time evolution operators is to implement the target $U$ itself on at-least $L$-qubit quantum systems for cost evaluation.
Our protocol can avoid this problem by compiling with smaller quantum systems with the size $\tilde{L}$, as large as the approximate causal cone by the LR bound, as discussed in Sec. \ref{Sec:Local_compilation}, 

\subsection{Lieb-Robinson bound}\label{Subsec:LiebRobinsonBound}
Lieb-Robinson (LR) bound dictates that any local observable cannot spread out faster than a certain finite velocity (called Lieb-Robinson velocity) under a local Hamiltonian \cite{Lieb1972-uo}. This can be interpreted as the emergence of approximate causal cones in quantum mechanics.

Let us describe it more precisely. We focus on a local Hamiltonian on a lattice $\Lambda$, given by
\begin{equation}\label{Eq:Local_Hamiltonian}
    H = \sum_{X \subseteq \Lambda} h_X,
\end{equation}
where $h_X$ denotes a term nontrivially acting on a domain $X \subseteq \Lambda$.
Let $\|\cdot\|$ denote the operator norm.
Here, we assume
\begin{enumerate}
    \item (Extensiveness) Local energy scale at every site is bounded by a finite value $g$;
    \begin{equation}\label{Eq:Extensiveness}
        \sum_{X; X \ni j} \| h_X \| \leq g, \quad \text{for any $j \in \Lambda$}.
    \end{equation}
    
    \item (Locality of interactions)
    At-most $k$-body interactions are involved with $k=O(1)$;
    \begin{equation}\label{Eq:Locality}
        h_X = 0, \quad \text{if} \quad |X| > k.
    \end{equation}
    
    \item (Range of interactions) Interactions are finite-ranged with distance $d_H = O(1)$;
    \begin{equation}\label{Eq:Finite_range}
        h_X = 0, \quad \text{if $\,^\exists j, j^\prime \in X$ s.t. $\mr{dist}(j,j^\prime) > d_H$}.
    \end{equation}
\end{enumerate}
Let us consider local observables $O_j$ and $O_{j^\prime}$ acting on $j$ and $j^\prime$ respectively, and assume that they are normalized as $\|O_j\|=\|O_{j^\prime}\|=1$. Then, the inequality, 
\begin{eqnarray}
    \| [U(\tau)^\dagger O_j U(\tau), O_{j^\prime}] \| &\leq& C e^{ - (\mr{dist}(j,j^\prime) - v \tau)/\xi }, \label{Eq:Lieb_Robinson_Bound} \\
    U(\tau) &=& e^{-i H \tau},
\end{eqnarray}
holds for a fixed time $\tau$. Here, the constant velocity $v$ and the constant length $\xi$ are determined only by the extensiveness $g$, the locality $k$, and the range $d_H$, while the constant $C$ depends on $\tau$ in addition ($C$ typically increases linearly in $\tau$ \cite{Lieb1972-uo}).

This suggests that $U(\tau)^\dagger O_j U(\tau)$ approximately acts on the domain inside the approximate causal cone $\{ j^\prime \in \Lambda \, | \, \mr{dist}(j,j^\prime) \leq v\tau \}$, and that the components outside of it are exponentially suppressed in the distance from $j$. 
As a result, it can be expected that $U(\tau)^\dagger O_j U(\tau)$ is well reproduced by the local Hamiltonian inside the approximate causal cone. 
Let $H^{(L^\prime,j)}$ denote the local Hamiltonian composed of $h_X$ whose support $X$ has distance from $j$ smaller than $L^\prime/2$ [See Eqs. (\ref{Eq:Restricted_lattice}) and (\ref{Eq:Restricted_Hamiltonian}) for the exact definition]. In fact, we can derive the following inequality from the LR bound (See Ref. \cite{Else2020-tr} and Appendix \ref{Asec:Extension});
\begin{equation}\label{Eq:Def_epsilon_LR1}
\| U(\tau)^\dagger O_j U(\tau) - e^{i H^{(L^\prime,j)} \tau} O_j e^{-i H^{(L^\prime,j)} \tau}\| \leq \varepsilon_\mr{LR},
\end{equation}
\begin{equation}\label{Eq:Def_epsilon_LR2}
\varepsilon_\mr{LR} = C^\prime \int_{L^\prime/2-d_H}^\infty e^{-(x-v\tau)/\xi} dx = e^{-O(l_0/\xi)},
\end{equation}
where $l_0$ is defined by $L^\prime = 2(l_0+d_H+v\tau)$. The integration comes from the summation all over the lattice out of the approximate causal cone. 

This relation enables us to approximate the local cost function (\ref{Eq:Cost_LHST}) for a large size $L$ by that for the smaller size $L^\prime$ with an arbitrarily small error $e^{-O(l_0/\xi)}$ when $L^\prime$ is sufficiently large compared to $v\tau$.

\section{Approximation of local cost functions by Lieb-Robinson bound}\label{Sec:Scalablity_of_cost}

\begin{figure}[t]
    \includegraphics[height=4.2cm, width=8.5cm]{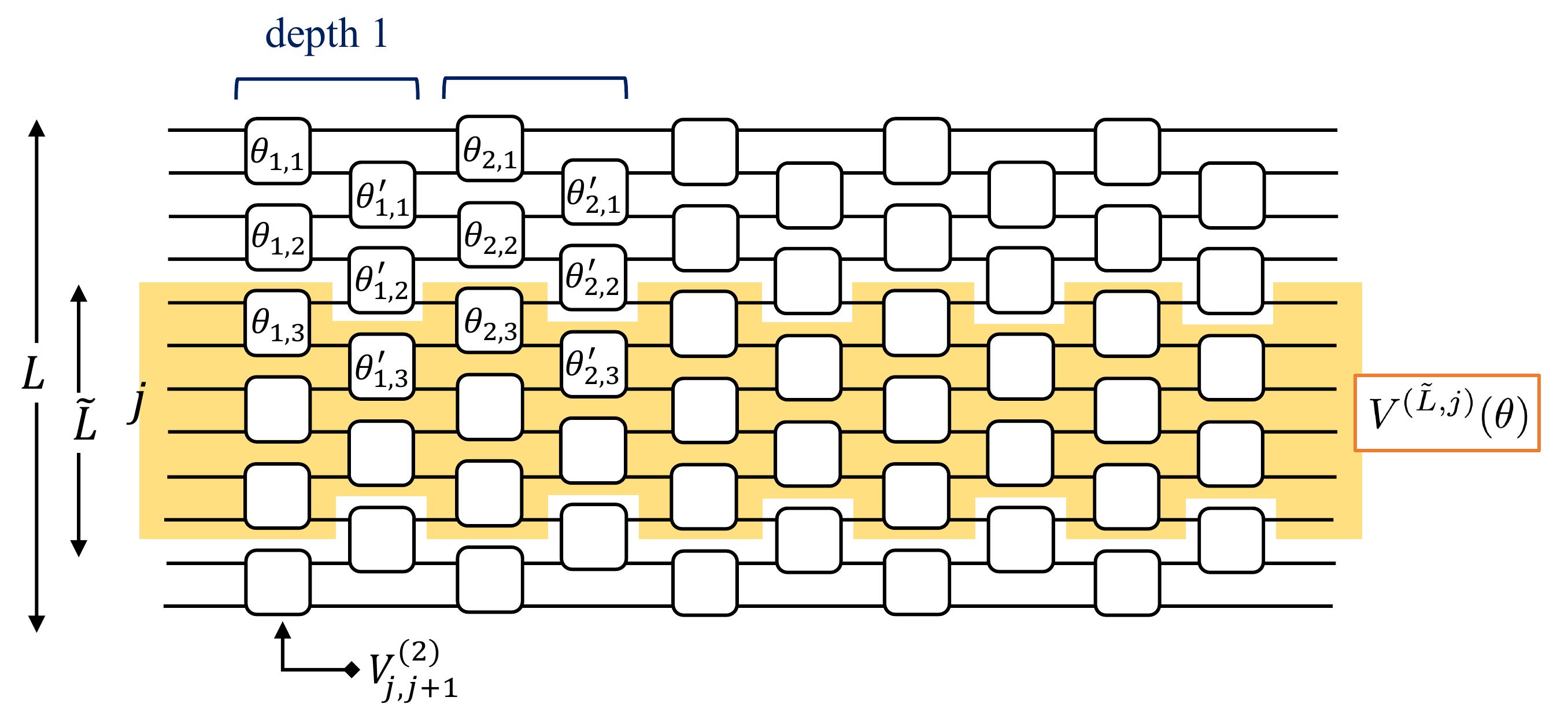}
    \caption{The brickwork-structured ansatz $V^{(L)}(\theta)$, defined by Eq. (\ref{Eq:Ansatz}). For translationally-invariant systems, we choose the variational parameter set $\theta = \{ \theta_{i,k}, \theta_{i,k}^\prime \}_{i,k}$ so that $\theta_{i,k}$ and $\theta_{i,k}^\prime$ respectively become independent of the position $k$. The yellow region represents the $j$-centered $\tilde{L}$-size domain $\Lambda^{(\tilde{L},j)}$, which is utilized for composing of the restricted ansatz $V^{(\tilde{L},j)}(\theta)$ based on Eq. (\ref{Eq:restricted_ansatz}).} 
    \label{fig:ansatz}
\end{figure}

In this section, we provide the first main result, where we compose of the subsystem cost functions and show their scaling property by the LR bound. 
The subsystem cost functions are obtained by the restriction of systems to smaller subsystems for the local cost function $C_\mr{LHST}$.
We clarify the approximate causal cone from the LR bound and the exact causal cone from the ansatz in the local cost functions.
They lead to two formulas, which are respectively raised as Propositions \ref{Prop:Result1} and \ref{Prop:Result2} below.
As a result, we obtain how the error between the subsystem cost functions and $C_\mr{LHST}$ scales in the subsystem size $\tilde{L}$, and validate the approximation of $C_\mr{LHST}$ by the subsystem cost functions with proper $\tilde{L}$.
As we will see in Sec. \ref{Sec:Local_compilation}, these results enables the LVQC protocol for the whole-system Hamiltonian dynamics.

First of all, we specify the setup and the notation. We consider a local and extensive Hamiltonian with finite-ranged interactions, $H$, on a lattice $\Lambda$ [See Eqs. (\ref{Eq:Extensiveness})-(\ref{Eq:Finite_range})]. 
Throughout the main text, we focus on an $L$-qubit one-dimensional system, as $\Lambda=\{1,2,\hdots, L \}$, but the extension to other cases is straightforward (See Appendix \ref{Asec:Extension}). 
We explicitly write the system size $L$ like $H^{(L)}$, and consider the target time evolution operator $U^{(L)}=\exp (-i H^{(L)} \tau)$.
For simplicity, we employ a brickwork-structured ansatz with the depth $d$ in the form of
\begin{equation}\label{Eq:Ansatz}
    V^{(L)} (\theta) = \prod_{i=1}^d \left[ \left(\prod_k V^{(2)}_{2k,2k+1}(\theta_{i,k})\right)\left(\prod_k V^{(2)}_{2k-1,2k}(\theta_{i,k}^\prime)\right)\right],
\end{equation}
as described in Fig. \ref{fig:ansatz}. Here, $V^{(2)}_{j,j^\prime}$ represents an arbitrary parametrized two-qubit gate on the neighboring sites $j$ and $j^\prime$, and the parameter set $\theta$ is composed of $\{ \theta_{i,k} \}_{i,k}$ and $\{ \theta_{i,k}^\prime \}_{i,k}$. 

Now, we derive two rigorous relations on the local cost function for each $j$-th site, $C_\mr{LHST}^{(j)}(U^{(L)}, V^{(L)})$, using the approximate causal cone from the LR bound and the exact causal cone from the locality of ansatz.
The first one, coming from the LR bound, validates the evaluation of the cost function with a local Hamiltonian acting only on qubits around $j$-th cite. 
To be precise, when we define the $j$-centered $L^\prime$-size domain $\Lambda^{(L^\prime,j)}$ and the restricted Hamiltonian $H^{(L^\prime,j)}$ by
\begin{eqnarray}
    \Lambda^{(L^\prime,j)} &=& \{ j^\prime \in \Lambda \, | \, |j-j^\prime| \leq L^\prime/2 \}, \label{Eq:Restricted_lattice} \\
    H^{(L^\prime,j)} &=& \sum_{X; X \subseteq \Lambda^{(L^\prime,j)}} h_X, \label{Eq:Restricted_Hamiltonian}
\end{eqnarray}
for the $L$-qubit Hamiltonian $H^{(L)}=\sum_{X; X \subseteq \Lambda} h_X$, they are related to the local cost functions $C_\mr{LHST}^{(j)}(U^{(L)}, V^{(L)})$ by the following proposition.

\begin{figure*}
\begin{center}
    \includegraphics[height=4cm, width=18cm]{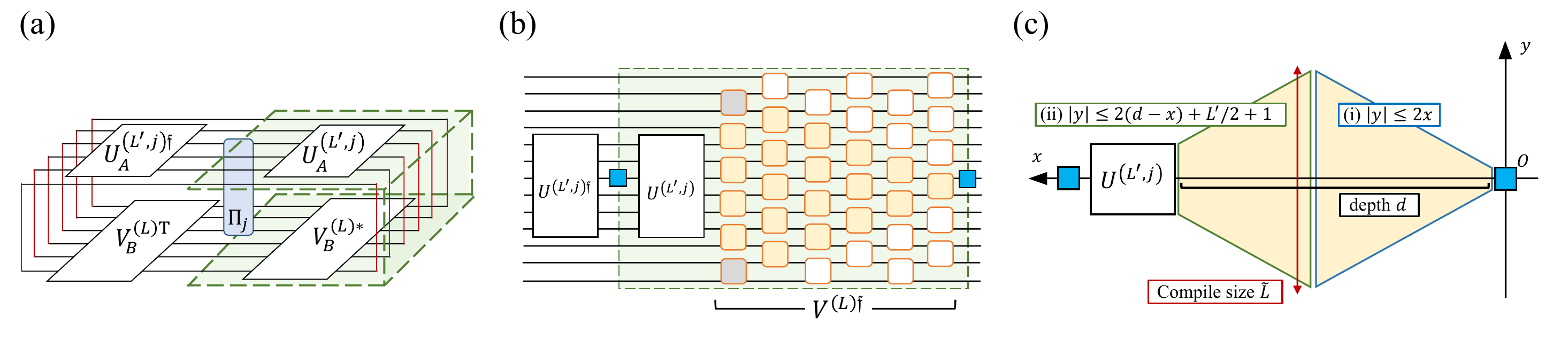}
    \caption{(a) Diagrammatic description of $\mr{Tr}(\Pi_j \rho_{AB}(U^{(L^\prime,j)},V^{(L)}))$. This gives an approximate upper bound of the local cost function $C_\mr{LHST}^{(j)}(U^{(L)},V^{(L)})$ via Proposition \ref{Prop:Result1}. (b) A part of gates composing $\mr{Tr}(\Pi_j \rho_{AB}(U^{(L^\prime,j)},V^{(L)}))$, designated by the green region in (a). Only the orange two-qubit gates in $V^{(L)\dagger}$ are active while the other gray and white two-qubit gates vanish due to their positions out of the causal cones. (c) Schematic picture of the active region in the ansatz $V^{(L)}$. All the active two-qubit gates are included in the yellow domain. Its height determines the proper compilation size $\tilde{L}$. } 
    \label{fig:proof}
 \end{center}
 \end{figure*}
 
\begin{proposition}\label{Prop:Result1}
Let the restriction size $L^\prime$ be chosen by
\begin{equation}
    L^\prime = 2(l_0+d_H+v\tau),
\end{equation}
with a tunable parameter $l_0$, the range of the Hamiltonian $d_H$, and the LR velocity $v$.
Then, the time evolution operator under the restricted Hamiltonian, defined by
\begin{equation}\label{Eq:restricted_U}
    U^{(L^\prime,j)}=e^{-i H^{(L^\prime,j)} \tau} \otimes I_{\Lambda \backslash \Lambda_{L^\prime,j}},
\end{equation}
provides the following inequality,
\begin{equation}\label{Eq:Result1}
C_{\mr{LHST}}^{(j)} (U^{(L)}, V^{(L)}) \leq 
 C_{\mr{LHST}}^{(j)} (U^{(L^\prime,j)}, V^{(L)}) + \frac{3}{4} \varepsilon_{\mr{LR}}.
\end{equation}
Here, the term $\varepsilon_\mr{LR}$ is defined by Eqs. (\ref{Eq:Def_epsilon_LR1}) and (\ref{Eq:Def_epsilon_LR2}), and it is exponentially small in the tunable parameter $l_0$ as $\varepsilon_\mr{LR}=e^{-O(l_0/\xi)}$.
\end{proposition}

\textit{Proof.}--- From the definition Eq. (\ref{Eq:Cost_LHST_j}), we obtain
\begin{eqnarray}
    && |C_{\mr{LHST}}^{(j)} (U^{(L)}, V^{(L)})- C_{\mr{LHST}}^{(j)} (U^{(L^\prime,j)}, V^{(L)})| \nonumber \\
    &\quad&  = |\mr{Tr}[\Pi_j \{ \rho_{AB}(U^{(L)},V^{(L)}) - \rho_{AB}(U^{(L^\prime,j)},V^{(L)}) \}]| \nonumber \\
    &\quad& = |\braket{\Phi_+ |(U_A^{(L)}\otimes V^{(L)\ast}_B)^\dagger \Pi_j (U_A^{(L)}\otimes V^{(L)\ast}_B) |\Phi_+}_{AB} \nonumber \\
    &\quad& \quad - \braket{\Phi_+ |(U_A^{(L^\prime,j)}\otimes V^{(L)\ast}_B)^\dagger \Pi_j (U_A^{(L^\prime,j)}\otimes V^{(L)\ast}_B) |\Phi_+}_{AB} |. \nonumber \\
    && \label{Eq:Proof1_A}
\end{eqnarray}
Considering that the projection to the Bell state is expanded by
\begin{eqnarray}
    \Pi_j &=& (\ket{\Phi_+}\bra{\Phi_+})_{A_j B_j} \nonumber \\
    &=& \frac{1}{4} (I_{A_j B_j}+X_{A_j}X_{B_j}-Y_{A_j}Y_{B_j}+Z_{A_j}Z_{B_j}), \label{Eq:Pi_j_decompose}
\end{eqnarray}
the right hand side of Eq. (\ref{Eq:Proof1_A}) is bounded by
\begin{eqnarray}
    && \frac{1}{4} \sum_{O=X,Y,Z} \| U^{(L)\dagger}_A O_{A_j} U^{(L)}_A - U^{(L^\prime,j)\dagger}_A O_{A_j}U^{(L^\prime,j)}_A \| \nonumber \\
    &\qquad& \qquad \qquad  \qquad  \times \| V^{(L) \mr{T}}_B O_{B_j} V^{(L)\ast}_B \| \braket{\Phi_+|\Phi_+}_{AB} \nonumber \\
    &&\leq \frac{3}{4} \varepsilon_\mr{LR}.
\end{eqnarray}
The above inequality comes from Eq. (\ref{Eq:Def_epsilon_LR1}), the LR bound for the local observable. Finally, we obtain the relation
\begin{equation}
     |C_{\mr{LHST}}^{(j)} (U^{(L)}, V^{(L)})- C_{\mr{LHST}}^{(j)} (U^{(L^\prime,j)}, V^{(L)})| \leq \frac{3}{4} \varepsilon_\mr{LR},
\end{equation}
which implies the inequality Eq. (\ref{Eq:Result1}). \hfill $\quad \square$

This proposition says that the restriction of the Hamiltonian to a smaller region hardly alters the local cost functions.
The difference is bounded by the LR bound error $\varepsilon_{\mathrm{LR}}$. 
Equivalently, the diagram of Fig. \ref{fig:preliminary} (b), which gives $C_\mr{LHST}$, can be approximated by that of Fig. \ref{fig:proof} (a), which gives the restricted version.
We note that this proof relies only on the existence of the LR bound, and hence Proposition \ref{Prop:Result1} is valid also for generic locally-interacting systems in any dimension.
For one-dimensional systems with finite-ranged interactions, we have $\varepsilon_\mr{LR}=\exp(-O(l_0/\xi))$ with $L^\prime=2(l_0+d_H+v\tau)$ from Eq. (\ref{Eq:Def_epsilon_LR2}). 
Based on this proposition, we can accurately determine the upper bound of the local cost function $C_{\mr{LHST}}^{(j)} (U^{(L)}, V^{(L)})$ by evaluating $C_{\mr{LHST}}^{(j)} (U^{(L^\prime,j)}, V^{(L)})$.

At this stage, however, measurement of the cost functions require $2L$-qubit quantum devices or $L$-qubit quantum devices with sampling due to the existence of $V^{(L)}$.
To overcome this obstacle, we employ causal cones of the ansatz $V^{(L)}$ and show that $C_{\mr{LHST}}^{(j)} (U^{(L^\prime,j)}, V^{(L)})$ can be evaluated with smaller-size quantum devices without any approximation. 
For the $j$-centered $\tilde{L}$-site domain $\Lambda_{\tilde{L},j}$, let us define a restricted ansatz $ V^{(\tilde{L},j)}(\theta)$ by
\begin{eqnarray}
    && V^{(\tilde{L},j)}(\theta) = \nonumber \\
    && \prod_{i=1}^d \left[ \left(\prod_k \!^{(\tilde{L},j)} V^{(2)}_{2k,2k+1}(\theta_{i,k})\right)\left(\prod_k \!^{(\tilde{L},j)} V^{(2)}_{2k-1,2k}(\theta_{i,k}^\prime)\right)\right], \nonumber \\
    && \label{Eq:restricted_ansatz}
\end{eqnarray}
from the $d$-depth ansatz $V^{(L)}(\theta)$ of Eq. (\ref{Eq:Ansatz}). Here, the symbols $\Pi_k^{(\tilde{L},j)}$ represent the product over $k$ such that the support of $V^{(2)}_{2k,2k+1}(\theta_{i,k})$ (for the first one) or $V^{(2)}_{2k-1,2k}(\theta_{i,k})$ (for the second one) is included in the domain $\Lambda_{\tilde{L},j}$ (See Fig. \ref{fig:ansatz}). Then, we obtain the following proposition.

\begin{proposition}\label{Prop:Result2}

We consider the same situation as that of Proposition \ref{Prop:Result1}. 
We assume $4d \geq L'$ for the $d$-depth $L$-site ansatz $V^{(L)}(\theta)$, and rewrite the depth as $d=L^\prime/4 + d^\prime$ ($d^\prime \geq 0$ is chosen so that $d$ becomes an integer).
For $\tilde{L}$ satisfying $\tilde{L} \geq L^\prime + 2d^\prime + 1$, where the right hand side represents the size of the approximate causal cones, the $d$-depth $\tilde{L}$-site ansatz $V^{(\tilde{L},j)}(\theta)$ satisfies the following equality;
\begin{equation}\label{Eq:Result2}
 C_{\mr{LHST}}^{(j)} (U^{(L^\prime,j)}, V^{(L)}) =  C_{\mr{LHST}}^{(j)} (\tilde{U}^{(L^\prime,j)}, V^{(\tilde{L},j)}).
\end{equation}
Here, $\tilde{U}^{(L^\prime,j)}$ represents the restriction of $U^{(L^\prime,j)}$ to the domain $\Lambda_{\tilde{L},j}$, which is given by
\begin{equation}\label{Eq:restricted_U_tilde}
    \tilde{U}^{(L^\prime,j)} = e^{-i H^{(L^\prime,j)} \tau} \otimes I_{\Lambda_{\tilde{L},j} \backslash \Lambda_{L^\prime,j}}.
\end{equation}

\end{proposition}
\textit{Remark.}--- The assumption $4d\geq L^\prime$ is not essentially required for proving this proposition. It rather serves as a guideline to construct the ansatz $V^{(L)}$. When we employ the brickwork-structured ansatz given by Eq. (\ref{Eq:Ansatz}), a local observable acting on a single qubit generally spreads to $4d$-qubit operators. Hence, we should use $d$ such that  $4d\geq L^\prime$ to capture the correlation within the LR bound and thereby accurately approximate the time evolution. It is straight-forward to generalize the above proposition to smaller $d$ with a slight modification of $\tilde{L}$.

\textit{Proof.}--- We employ the causal cones of quantum circuits here. Let us focus on $\mr{Tr} [\Pi_j \rho_{AB}(U^{L^\prime,j}, V^{(L)})]$, which can be schematically depicted by Fig. \ref{fig:proof} (a). 
To visualize the causal cone, we pick up a part of the circuit belonging to the right half in the figure (the light-green region), which results in Fig. \ref{fig:proof} (b). 
The light-blue squares in Fig. \ref{fig:proof} (b) represent local operators on $j$-th sites composing $\Pi_j$, given by Eq. (\ref{Eq:Pi_j_decompose}).
We also note that $V_B^{(L)\ast}$ in Fig. \ref{fig:proof} (a) is translated into $V^{(L) \dagger}$ since its input and output are exchanged. 

Each local two-qubit gates in the ansatz $V^{(L) \dagger}$ can be classified into three groups by its effect on the local cost function. 
The first group is depicted by the white (non-painted) two-qubit gates in Fig. \ref{fig:proof} (b). Since these local gates and the corresponding ones in $V^{(L) \mr{T}}_B$ cancel each other by the contraction in the lower layer of Fig. \ref{fig:proof} (a), they do not affect $\mr{Tr} [\Pi_j \rho_{AB}(U^{(L^\prime,j)}, V^{(L)})]$. This cancellation is due to the locality of $\Pi_j$ and independent of $U_A^{(L^\prime,j)}$ appearing in the upper layer.
The second group, composed of the gray two-qubit gates, are also inactive, because they can be contracted to identity in the upper layer. In contrast to the first group, its cancellation originates from the size restriction of the Hamiltonian $H^{(L)}$ to $H^{(L^\prime)}$, validated by the LR bound.
The last one is composed of the yellow gates residing within the causal cones. Only these two-qubit gates are relevant for $\mr{Tr} [\Pi_j \rho_{AB}(U^{(L^\prime,j)}, V^{(L)})]$, which can be schematically depicted as Fig. \ref{fig:proof} (c).

Finally, we determine the proper compilation size $\tilde{L}$. 
The active region, composed of the two causal cones spreading from the left- and the right-side [See (i) and (ii) in Fig. \ref{fig:proof} (c)], is designated by
\begin{equation}
    |y| \leq \min \{ 2 (d-x) + L^\prime/2 +1, 2x \}, \quad  0 \leq x \leq d.
\end{equation}
When the compilation  size $\tilde{L}$ surpasses its height, that is, when
\begin{equation}
    \tilde{L} \geq \frac{L^\prime}{2} + 2d +1 = L^\prime + 2d^\prime +1,
\end{equation}
the restricted ansatz $V^{(\tilde{L},j)}(\theta)$ includes all the two-qubit gates in the active region. Therefore, we have
\begin{eqnarray}
    && \mr{Tr} [\Pi_j \rho_{AB}(U^{(L^\prime,j)}, V^{(L)})] \nonumber \\
    && = \mr{Tr} [\Pi_j \rho_{AB}(e^{-iH^{(L^\prime,j)}\tau} \otimes I_{\Lambda \backslash \Lambda_{L^\prime,j}}, V^{(\tilde{L},j)} \otimes I_{\Lambda \backslash \Lambda_{\tilde{L},j}})] \nonumber \\
    && = \mr{Tr} [\Pi_j \rho_{AB}(e^{-iH^{(L^\prime,j)}\tau} \otimes I_{\Lambda_{\tilde{L},j} \backslash \Lambda_{L^\prime,j}}, V^{(\tilde{L},j)}) ],
\end{eqnarray}
where we use the fact that the contraction over $\Lambda \backslash \Lambda_{\tilde{L},j}$ gives identity for the last equality. 
By using the definitions of the local cost function $C_\mr{LHST}^{(j)}$ and the restricted time evolution $\tilde{U}^{(L^\prime,j)}$ [See Eqs. (\ref{Eq:Cost_LHST_j}) and (\ref{Eq:restricted_U_tilde}) respectively], we complete the proof of Proposition \ref{Prop:Result2}. \hfill $\quad \square$

Let us define the subsystem cost function by $C_{\mr{LHST}}^{(j)} (\tilde{U}^{(L^\prime,j)}, V^{(\tilde{L},j)})$, which can be measured by a $2\tilde{L}$-qubit quantum device or a $\tilde{L}$-qubit quantum device with a Monte-Carlo sampling based on Lemma \ref{Thm:LqubitLHST}.
Propositions \ref{Prop:Result1} and \ref{Prop:Result2} yield that the local cost function $C_\mr{LHST}^{(j)}(U^{(L)},V^{(L)})$ can be approximated by the subsystem cost function, and they also dictate the scaling property of the subsystem cost function in the subsystem size $\tilde{L}$.
Importantly, $\tilde{L} \geq L^\prime + 2d^\prime +1 = 2 (l_0 + d_H + v\tau ) + 2d^\prime + 1$ can be independent of the whole-system size $L$ and significantly smaller than $L$.
We note that the coefficient of the depth $d$ in $\tilde{L}$ comes from the brickwork structure of the ansatz $V^{(L)}$. We can obtain the same result for any other ansatz with changing the coefficient in $\tilde{L}$ as long as it is local.

\section{Local variational quantum compilation of a large-scale Hamiltonian dynamics}\label{Sec:Local_compilation}

In this section, we formulate the local variational quantum compilation (LVQC) of a large-scale Hamiltonian dynamics as the second main result. In our protocol, we construct an approximate time evolution operator for the large size $L$ by optimizing the cost functions defined on the smaller size $\tilde{L}$.
Based on Propositions \ref{Prop:Result1} and \ref{Prop:Result2}, we provide two different formulations for translationally-invariant cases (Sec. \ref{Subsec:Local_compilation_PBC}) and generic cases (Sec. \ref{Subsec:local_compilation_generic}).

\subsection{Local compilation for translationally-invariant systems}\label{Subsec:Local_compilation_PBC}
We first deal with translationally-invariant cases under periodic boundary conditions (PBC). Throughout this section, we denote such a translationally invariant Hamiltonian and its time evolution operator for the size $L$ as $H^{(L)}_\mr{PBC}$ and $U^{(L)}_\mr{PBC}$.
Then, it is reasonable to impose translation invariance and PBC also on the ansatz, denoted by $V^{(L)}_\mr{PBC}$. To be precise, we assume that the variational parameter set $\theta = \{ \theta_{i,k}, \theta^\prime_{i,k} \}_{i,k}$ is independent of the position $k$. The number of parameters in $V^{(L)}_\mr{PBC}(\theta)$ depends only on the depth $d$. Based on Propositions \ref{Prop:Result1} and \ref{Prop:Result2}, we can derive the following theorem which also shows the protocol of LVQC.

\begin{theorem}\label{Thm:local_compilation_PBC}
We define the local-compilation cost function by,
\begin{eqnarray}
    &&C_\alpha^{(\tilde{L})} (\theta) = \nonumber \\
    && \alpha C_\mr{HST}(U^{(\tilde{L})}_\mr{PBC}, V^{(\tilde{L})}_\mr{PBC}(\theta)) + (1-\alpha)  C_\mr{LHST}(U^{(\tilde{L})}_\mr{PBC}, V^{(\tilde{L})}_\mr{PBC}(\theta)), \nonumber \\
    &&    
\end{eqnarray}
for a certain $\alpha \in [0,1]$, which is defined on an $\tilde{L}$-size translationally-invariant systems under PBC.
Assume that, after the minimization of $C_\alpha^{(\tilde{L})} (\theta)$, the optimal parameter set $\theta_\mr{opt}$ gives the upper bound of the local and global cost functions as
\begin{eqnarray}
    && C_\mr{LHST}(U_\mr{PBC}^{(\tilde{L})}, V_\mr{PBC}^{(\tilde{L})}(\theta_\mr{opt})) < \varepsilon_\mr{LHST}, \label{Eq:thm_PBC_assume_LHST} \\
    && C_\mr{HST}(U_\mr{PBC}^{(\tilde{L})}, V_\mr{PBC}^{(\tilde{L})}(\theta_\mr{opt})) < \varepsilon_\mr{HST}. \label{Eq:thm_PBC_assume_HST}
\end{eqnarray}
When we choose the smallest even number larger than $2(l_0+d_H+v\tau) +2d^\prime+1$ as the compilation size $\tilde{L}$, the time evolution operator for an $L$-qubit system ($L \geq \tilde{L}$) is approximated as
\begin{eqnarray}
&& C_\mr{LHST} (U_\mr{PBC}^{(L)}, V_\mr{PBC}^{(L)} (\theta_\mr{opt})) \leq  \varepsilon_\mr{LHST} + \frac{3}{2} \varepsilon_\mr{LR}, \label{Eq:thm_PBC_result_LHST} \\
&& C_\mr{HST} (U_\mr{PBC}^{(L)}, V_\mr{PBC}^{(L)} (\theta_\mr{opt})) \leq L \left( \varepsilon_\mr{HST} + \frac{3}{2} \varepsilon_\mr{LR} \right), \label{Eq:thm_PBC_result_HST}
\end{eqnarray}
with the usage of the same parameter set $\theta_\mr{opt}$.
\end{theorem}

\begin{figure}[t]
    \includegraphics[height=3.9cm, width=8.5cm]{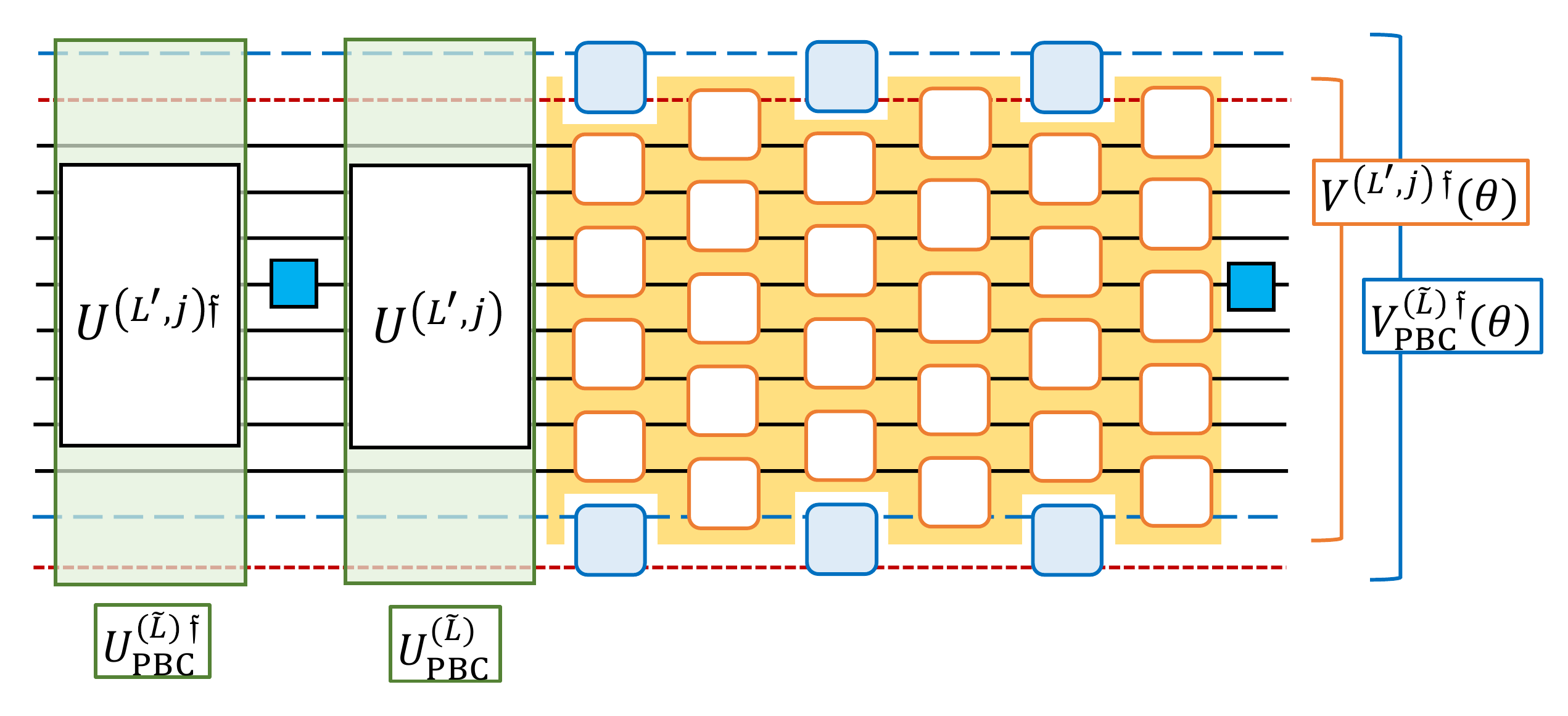}
    \caption{Schematic picture of Fig. \ref{fig:proof} (b) for translationally-invariant systems under PBC. The blue and red solid lines respectively represent identical sites.} 
    \label{fig:pbc_circuit}
\end{figure}

\textit{Proof.}--- We first derive Eq. (\ref{Eq:thm_PBC_result_LHST}) from Eq. (\ref{Eq:thm_PBC_assume_LHST}). 
We combine translation symmetry with the scaling property of the subsystem cost functions, represented by Propositions \ref{Prop:Result1} and \ref{Prop:Result2}. As a result, we obtain the relation for any $j$,
\begin{eqnarray}
 C_\mr{LHST} (U_\mr{PBC}^{(L)}, V_\mr{PBC}^{(L)}) &=& C_\mr{LHST}^{(j)} (U_\mr{PBC}^{(L)}, V_\mr{PBC}^{(L)}) \nonumber \\
 &\leq& C_\mr{LHST}^{(j)} (\tilde{U}^{(L^\prime,j)}, V^{(\tilde{L},j)}) + \frac{3}{4} \varepsilon_\mr{LR}. \nonumber \\
 && \label{Eq:proof_PBC_1}
\end{eqnarray}
Here, $\tilde{U}^{(L^\prime,j)}$ and $V^{(\tilde{L},j)}$ are constructed from $U_\mr{PBC}^{(L)}$ and $V_\mr{PBC}^{(L)}$ by the restriction to $L^\prime$- and $\tilde{L}$-size systems respectively [See Eqs. (\ref{Eq:restricted_U_tilde}) and (\ref{Eq:restricted_ansatz})]. 
They have open boundary condition (OBC) as illustrated in Fig. \ref{fig:pbc_circuit}, and therefore do not straight-forwardly relate to $U_\mr{PBC}^{(\tilde{L})}$ and $V_\mr{PBC}^{(\tilde{L})}$.
To recover the PBC, we take the following strategy. Figure \ref{fig:pbc_circuit} gives a schematic picture of a part of gates composing $\mr{Tr}[\Pi_j \rho_{AB}(\tilde{U}^{L^\prime,j},V^{(\tilde{L},j)})]$, similar to Fig. \ref{fig:proof} (b).
First, we add two-qubit gates $V^{(2)}_{\tilde{L},1}$ to each layer of the restricted ansatz $V^{(\tilde{L},j)}$, represented by the light-blue squares at the boundaries in Fig. \ref{fig:pbc_circuit}.
When the parameter set of each $V^{(2)}_{\tilde{L},1}$ is same as that of the two-qubit gate in the same layer, it reproduces the translationally-invariant ansatz under PBC, $V^{(\tilde{L})}_\mr{PBC}$.
Since local gates outside of the causal cones does not alter the local cost function at all, we obtain the following relation;
\begin{equation}\label{Eq:proof_PBC_2}
C_\mr{LHST} (U_\mr{PBC}^{(L)}, V_\mr{PBC}^{(L)}) \leq C_\mr{LHST}^{(j)} (\tilde{U}^{(L^\prime,j)}, V^{(\tilde{L})}_\mr{PBC}) + \frac{3}{4} \varepsilon_\mr{LR}.
\end{equation}
We also recover the PBC of the target unitary $\tilde{U}^{(L^\prime,j)}$. Let us consider the two Hamiltonians $H^{(\tilde{L})}_\mr{PBC}$ and $H^{(L^\prime,j)}$, which respectively provide the time evolution operators $U^{(\tilde{L})}_\mr{PBC}$ and $\tilde{U}^{(L^\prime,j)}$.
Since the Hamiltonian $H^{(L^\prime,j)}$ becomes the restriction of $H^{(\tilde{L})}_\mr{PBC}$ from the domain $\Lambda_{\tilde{L},j}$ to the one $\Lambda_{L^\prime,j}$, we can again employ the inequality Eq. (\ref{Eq:Def_epsilon_LR1}) brought by the LR bound,
\begin{equation}
    \| U^{(\tilde{L}) \dagger}_\mr{PBC} O_j U^{(\tilde{L})}_\mr{PBC} - \tilde{U}^{(L^\prime,j) \dagger} O_j \tilde{U}^{(L^\prime,j)} \| \leq \varepsilon_\mr{LR},
\end{equation}
for any local normalized observable at a $j$-th site, $O_j$.
This implies that we can apply Propositions \ref{Prop:Result1} and \ref{Prop:Result2} with substituting $U^{(\tilde{L})}_\mr{PBC}$ for $U^{(L)}$, which results in
\begin{eqnarray}
    C_\mr{LHST}^{(j)} (\tilde{U}^{(L^\prime,j)}, V^{(\tilde{L})}_\mr{PBC}) &\leq& C_\mr{LHST}^{(j)} (U^{(\tilde{L})}_\mr{PBC}, V^{(\tilde{L})}_\mr{PBC}) + \frac{3}{4} \varepsilon_\mr{LR} \nonumber \\
    &<& \varepsilon_\mr{LHST} + \frac{3}{4} \varepsilon_\mr{LR}.
\end{eqnarray}
Combining this inequality with Eq. (\ref{Eq:proof_PBC_2}), we arrive at the relation of $C_\mr{LHST}$, given by Eq. (\ref{Eq:thm_PBC_result_LHST}).

Next, we derive Eq. (\ref{Eq:thm_PBC_result_HST}), which gives an upper bound of the global cost function $C_\mr{HST}$.
We employ the following inequality \cite{Khatri2019-gj},
\begin{equation}\label{Eq:relation_HST_LHST}
    C_\mr{LHST} (U,V) \leq C_\mr{HST} (U, V) \leq |\Lambda| C_\mr{LHST} (U,V),
\end{equation}
when two unitary gates $U$ and $V$ are defined on a lattice $\Lambda$. 
Under the assumption of Eq. (\ref{Eq:thm_PBC_assume_HST}), we have $C_\mr{LHST}(U_\mr{PBC}^{(\tilde{L})}, V_\mr{PBC}^{(\tilde{L})}(\theta_\mr{opt})) < \varepsilon_\mr{HST}$ from the first inequality in Eq. (\ref{Eq:relation_HST_LHST}). 
Using the above result for the local cost function $C_\mr{LHST}$, Eq. (\ref{Eq:thm_PBC_result_LHST}), we obtain $C_\mr{LHST} (U_\mr{PBC}^{(L)}, V_\mr{PBC}^{(L)} (\theta_\mr{opt})) \leq  \varepsilon_\mr{HST} + \frac{3}{2} \varepsilon_\mr{LR}$. 
Finally, considering $|\Lambda| = L$ for a one-dimensional system, the second inequality in Eq. (\ref{Eq:relation_HST_LHST}) implies Eq. (\ref{Eq:thm_PBC_result_HST}). \hfill $\quad \square$

This theorem tells us that the optimal parameter set $\theta_\mr{opt}$ for the $\tilde{L}$-size local-compilation cost function can be directly employed to construct the approximate larger-scale time evolution by $U^{(L)}_\mr{PBC} \simeq V^{(L)}_\mr{PBC}(\theta_\mr{opt})$.
Its accuracy can be guaranteed by Eq. (\ref{Eq:thm_PBC_result_LHST}) or Eq. (\ref{Eq:thm_PBC_result_HST}).
The error consists of two parts: the first terms, $\varepsilon_\mr{LHST}$ and $\varepsilon_\mr{HST}$, are due to a limited expressive power of the ansatz $V^{(\tilde{L})}_\mr{PBC}$; the second term, $\varepsilon_\mr{LR}$ is the intrinsic error induced by this LVQC protocol. 
They can be improved by using more expresive ansatz and using larger compilation size $\tilde{L}$, respectively.

Now, we discuss what compilation size should be used to achieve an accuracy of $O(\varepsilon)$ for a quantity of interest.
When we focus on some local observables under the approximate time evolution $V^{(L)}_\mr{PBC}(\theta_\mr{opt})$, the local cost function $C_\mr{LHST}$ plays a significant role since it guarantees the local equivalence with $U^{(L)}_\mr{PBC}$ by Eq. (\ref{Eq:LHST_zero_condition}).
To be more precise, $C_\mr{LHST}=O(\varepsilon)$ implies additive error $O(\varepsilon)$ in the expectation values of local observables.
We wish to choose the compilation size $\tilde{L}= 2 \lceil l_0+d_H+v\tau+d^\prime + 1/2 \rceil$ so that $\varepsilon_\mr{LR} =  e^{-O(l_0/\xi)}$ can be neglected.
Therefore, in this case, $\tilde{L}$ can be taken as $ O(\xi\log(1/\epsilon))+2d_H+2v\tau+2d^\prime$, which is independent of the whole-system size $L$.

On the other hand, in the cases where we require the accuracy in terms of global observables, the average gate fidelity $\bar{F}$ has the operational meaning.
$1-\bar{F}=O(\varepsilon)$ implies an accuracy of $O(\varepsilon)$ in the expectation values of any observables.
When the $\tilde{L}$-size optimization is achieved as Eqs. (\ref{Eq:thm_PBC_assume_LHST}) and (\ref{Eq:thm_PBC_assume_HST}), the combination with Eqs. (\ref{Eq:operational_meaning_HST}) or (\ref{Eq:operational_meaning_LHST}) ensures its lower bound as
\begin{eqnarray}
\bar{F} (U_\mr{PBC}^{(L)}, V_\mr{PBC}^{(L)} (\theta_\mr{opt})) &\geq& 1 - \frac{2^{|\Lambda|}}{2^{|\Lambda|}+1} \cdot L \left( \varepsilon_\mr{LHST} + \frac{3}{2} \varepsilon_\mr{LR} \right) \nonumber \\
&\geq& 1 - \frac{2^{|\Lambda|}}{2^{|\Lambda|}+1} \cdot L \left( \varepsilon_\mr{HST} + \frac{3}{2} \varepsilon_\mr{LR} \right). \nonumber \\
&& \label{Eq:average_gate_fidelity_pbc}
\end{eqnarray}
Therefore, to achieve $1-\bar{F}=O(\varepsilon)$, we should choose the compilation size $\tilde{L}$ satisfying $L \varepsilon_\mr{LR} = e^{-O(l_0/\xi) + \log L} =O(\varepsilon)$, which results in $\tilde{L} = O(\xi\log{(1/\varepsilon)}+\xi\log{L}) + 2 d_H + 2 v\tau + 2d^\prime$.
Upon this choice of the compilation size, we should continue the optimization of $C_\alpha^{(\tilde{L})} (\theta)$ until $\varepsilon_\mr{LHST}$ or $\varepsilon_\mr{HST}$ becomes much smaller than $O(L^{-1})$, and then we can obtain preferable accuracy.

To summarize, our protocol starts with choosing a proper compilation size $\tilde{L}$.
$\tilde{L}$ should taken to be comparable to the approximate causal cone size by the LR bound, $2 (\xi+ d_H+v\tau+d^\prime)$, or a bit larger than it, depending on the desired error.
After minimizing the local-compilation cost function $C_\alpha^{(\tilde{L})} (\theta)$ which can be evaluated using classical simulator or quantum device with at least $\tilde{L}$ qubits, we can directly apply the optimal parameter set $\theta_\mr{opt}$ to obtain the approximate time evolution $U^{(L)}_\mr{PBC} \simeq V^{(L)}_\mr{PBC}(\theta_\mr{opt})$.
This reduction in the size makes NISQ devices or classical simulators employing some approximation (See Sec. \ref{Sec:Numerical}) suitable for the compilation.
LVQC can be employed for various purposes such as depth compression and calibration of $U^{(L)}$, without implementing the target $U^{(L)}$ itself but only with the one for smaller systems $U^{(\tilde{L})}$. This is clearly one of the largest advantages in our protocol.
In addition, by repeating the application of $ V(\theta_\mr{opt})$, we can approximately simulate the stroboscopic dynamics at $t=n\tau$ ($n \in \mathbb{N}$). Thus, LVQC with the size-scale $\tilde{L}$ and the time-scale $\tau$ can be applied to reproduce the dynamics of larger scales both in space and time. 
We summarize the results in Fig. \ref{fig:summary}.

\subsection{Local compilation for generic systems without translation-invariance}\label{Subsec:local_compilation_generic}
Here, we develop the LVQC protocol for one-dimensional finite-ranged systems without translation-invariance. The result does not essentially alter from translationally-invariant cases, but they have different cost functions.

We directly use Propositions \ref{Prop:Result1} and \ref{Prop:Result2} to derive the protocol. For the brickwork-structured ansatz $V^{(L)}(\theta)$ (not necessarily translationally-invariant), we define the local-compilation cost function for generic cases by
\begin{equation}\label{Eq:Local_cost_generic_case}
    C^{(\tilde{L})}(\theta) = \frac{1}{L} \sum_{j=1}^L C_\mr{LHST}^{(j)} (\tilde{U}^{(L^\prime,j)}, V^{(\tilde{L},j)}(\theta)),
\end{equation}
where we directly use the subsystem cost functions $C_\mr{LHST}^{(j)} (\tilde{U}^{(L^\prime,j)}, V^{(\tilde{L},j)}(\theta))$.
With the help of Propositions \ref{Prop:Result1} and \ref{Prop:Result2}, we immediately obtain
\begin{eqnarray}
    && |C_\mr{LHST}(U^{(L)},V^{(L)}) - C^{(\tilde{L})}(\theta) | \nonumber \\
    &\quad& \leq    \frac{1}{L} \sum_{j=1}^L |C_\mr{LHST}^{(j)}(U^{(L)},V^{(L)}) - C_\mr{LHST}^{(j)}(\tilde{U}^{(L^\prime,j)},V^{(\tilde{L},j)})| \nonumber \\
    &\quad& \leq \frac{3}{4} \varepsilon_\mr{LR}.
\end{eqnarray}
We also use the relation Eq. (\ref{Eq:relation_HST_LHST}), which results in
\begin{equation}
    C_\mr{HST}(U^{(L)},V^{(L)}) \leq L \left( C^{(\tilde{L})}(\theta) + \frac{3}{4} \varepsilon_\mr{LR} \right).
\end{equation}
Therefore, we obtain the following theorem, which designates the protocol for generic cases.
\begin{theorem}\label{Thm:Local_compilation_generic}
We variationally minimize the local-compilation cost function $C^{(\tilde{L})}(\theta)$. When the optimal parameter set $\theta_\mr{opt}$ gives $C^{\tilde{L}}(\theta_\mr{opt}) \leq \varepsilon_\mr{LHST}$, the cost functions for the size $L$ is bounded by
\begin{eqnarray}
    && C_\mr{LHST}(U^{(L)}, V^{(L)}(\theta_\mr{opt})) \leq \varepsilon_\mr{LHST} + \frac{3}{4} \varepsilon_\mr{LR}, \label{Eq:thm_generic_result_LHST} \\
    && C_\mr{HST}(U^{(L)}, V^{(L)}(\theta_\mr{opt})) \leq L \left( \varepsilon_\mr{LHST} + \frac{3}{4} \varepsilon_\mr{LR} \right). \label{Eq:thm_generic_result_HST}
\end{eqnarray}
The average gate fidelity is bounded from below as follows:
\begin{equation}\label{Eq:average_gate_fidelity_generic}
    \bar{F}(U^{(L)}, V^{(L)}(\theta_\mr{opt})) \geq 1- \frac{2^{|\Lambda|}}{2^{|\Lambda|+1}} \cdot L \left( \varepsilon_\mr{LHST} + \frac{3}{4} \varepsilon_\mr{LR} \right).
\end{equation}
\end{theorem}

\begin{figure*}
\begin{center}
    \includegraphics[height=5cm, width=18cm]{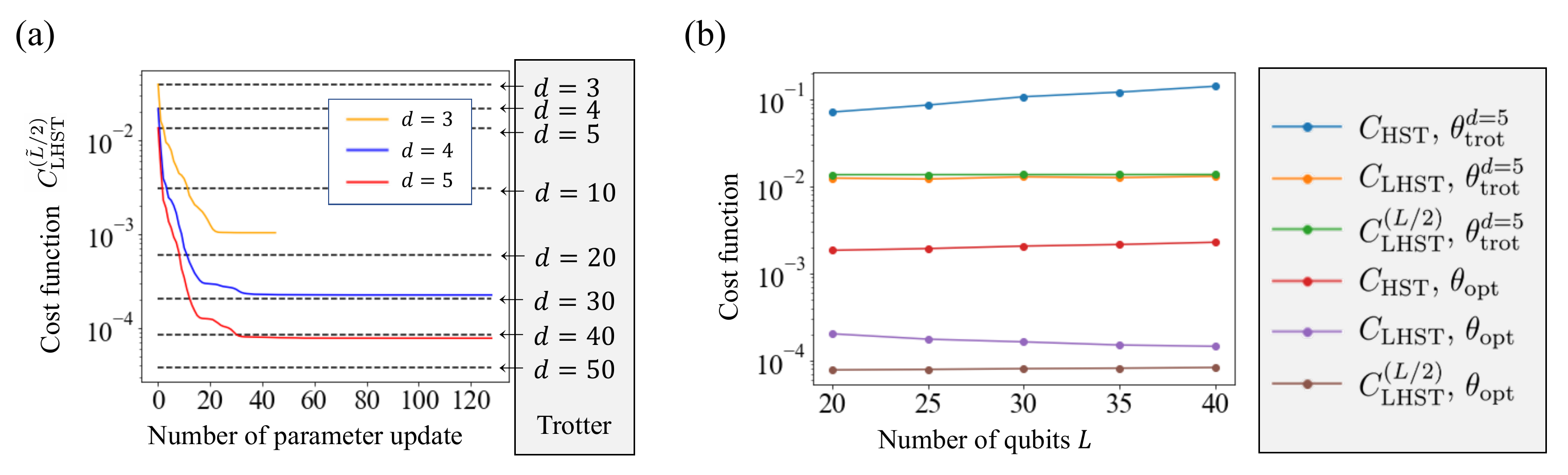}
    \caption{(a) The history of the cost function $C_\mr{LHST}^{(\tilde{L}/2)}(U^{(\tilde{L})}, V^{(\tilde{L})}(\theta))$ in the intermediate size $\tilde{L}=20$. The yellow, blue, and the red solid lines respectively represent the results for the depth of the ansatz $d=3,4,5$. The dashed lines represent the corresponding cost functions for the Trotter decomposition with various depth $d$. (b) Cost functions $C(U^{(L)},V^{(L)}(\theta))$ for increasing $L$. The ansatz $V^{(L)}(\theta_\mr{opt})$, where $\theta_\mr{opt}$ is obtained by the optimization using only $\tilde{L}=20$ qubits, well approximates $U^{(L)}$ compared to the Trotter decomposition with the same depth, $V^{(L)}(\theta_\mr{trot}^{d=5})$.} 
    \label{fig:numerical}
 \end{center}
 \end{figure*}
 
Based upon this theorem, we can perform the local compilation in a similar way to translationally-invariant systems, while the cost function is replaced by Eq. (\ref{Eq:Local_cost_generic_case}). 
We have the same compilation size $\tilde{L} = 2 \lceil l_0+d_H+v\tau+d^\prime+1/2 \rceil$ with $l_0$ such that $\varepsilon_\mr{LR}$ or $L \varepsilon_\mr{LR}$ becomes sufficiently small.
After the local optimization that achieves $\varepsilon_\mr{LHST} \ll 1$ or $L \varepsilon_\mr{LHST} \ll 1$, we use the optimal parameter set $\theta_\mr{opt}$ for the $L$-size time evolution operator as schematically shown in Fig. \ref{fig:summary}.

We also remark extension of our protocol to other generic cases. Our protocol relies only on the existence of the LR bound, given by Eq. (\ref{Eq:Def_epsilon_LR1}), and the locality of the ansatz.
Thus, the extension to higher-dimensional systems is straightforward, in which we change the form of $\varepsilon_\mr{LR}$ and replace the coefficient $L$ in Eqs. (\ref{Eq:thm_PBC_result_HST}) or (\ref{Eq:thm_generic_result_HST}) by $|\Lambda| \sim L^D$. 
We can also consider short-ranged, or long-ranged interactions since they respectively show an exponential or polynomial decay of the error $\varepsilon_\mr{LR}$ (Note that we require additional conditions when considering long-ranged interactions for the existence of the LR bound, as discussed in Appendix \ref{Asubsec:Extension_Long_range}). 
The compilation size $\tilde{L}$ increases at-most in $O(\log L)$ (for finite-ranged, short-ranged interactions in generic dimension) or in $O(L^\sigma)$ with $\sigma<1$ (long-ranged interactions in generic dimension).
We can expect significant reduction in the compilation size for a broad class of locally-interacting systems to compile large-scale time evolution operators. See Appendix \ref{Asec:Extension} for the detailed discussion.

\section{Numerical demonstration of depth compression}\label{Sec:Numerical}

Here, we numerically demonstrate LVQC, and in particular, we try to compress the depth of a large-scale time evolution operator by the compilation. 
For simplicity, we concentrate on one-dimensional systems and rely on classical simulation by time-evolving block decimation (TEBD), based on matrix product states (MPS) \cite{Vidal2003-eb,Vidal2004-oa,Hastings2009-hd,Schollwock2011-zj}.

We first introduce the model and the ansatz. We adopt an anti-ferromagnetic (AFM) Heisenberg model on a one-dimensional lattice, defined by
\begin{equation}
    H_\mr{AFM}^{(L)} = \sum_{j=1}^{L-1} (X_j X_{j+1} + Y_j Y_{j+1} + Z_j Z_{j+1}).
\end{equation}
We employ OBC to make it easier to simulate by MPS. 
The target of the depth compression is the time evolution operator $U^{(L)} = \exp (- i H_\mr{AFM}^{(L)} \tau)$ with a fixed time $\tau$.
On the other hand, we give the ansatz $V^{(L)}(\theta)$ by the brickwork-structured circuit under OBC, designated by Eq. (\ref{Eq:Ansatz}). We parameterize each of two-qubit gates in it by
\begin{eqnarray}
    && V_{j,j+1}^{(2)} (\eta, \zeta, \chi, \gamma, \phi)  = \nonumber \\ && \quad \left( \begin{array}{cccc}
    1 & 0 & 0 & 0 \\
    0 & e^{-i(\gamma+\zeta)}\cos \eta & -i e^{-i(\gamma-\chi)} \sin \eta & 0 \\
    0 & -i e^{-i(\gamma+\chi)}\sin \eta & e^{-i (\gamma - \zeta)} \cos \eta & 0 \\
    0 & 0 & 0 & e^{-i (2\gamma+\phi)}
    \end{array}
\right),\nonumber \\
&&
\end{eqnarray}
in the basis of $\{ \ket{00}, \ket{01}, \ket{10}, \ket{11} \}$, where $\eta$, $\zeta$, $\chi$, $\gamma$, and $\phi$ denote the variational parameters. 
This form is chosen so that $V_{j,j+1}^{(2)}$ can represent any two-qubit gate preserving the total $Z$-spin which is a symmetry of  $H_\mr{AFM}$ \cite{Arute2020-qr,Neill2021-jv}. 
Here, we expect that, when the system has more than tens of qubits, its boundaries hardly affect the results.
Reflecting this approximate translation symmetry, we employ a single parameter set $(\eta, \zeta, \chi, \gamma, \phi)$ within each layer. 
Upon this setup, the number of the independent variational parameters becomes $10d$ for the $d$-depth ansatz.

We examine whether we can approximate $U^{(L)}$ by the shallow-depth circuit $V^{(L)}$ as $U^{(L)} \simeq V^{(L)}(\theta_\mr{opt})$, with the optimal parameter set obtained in the smaller size $\tilde{L}$. 
Based on the approximate translation symmetry, we apply the protocol for translationally-invariant systems under PBC.
To be precise, based on Theorem \ref{Thm:local_compilation_PBC}, we minimize the local cost function $C_\mr{LHST}^{(\tilde{L}/2)}(U^{(\tilde{L})}, V^{(\tilde{L})}(\theta))$, which is expected to approximate the local-compilation cost function $C^{(\tilde{L})}_{\alpha=0}(\theta)$.
Then, with the optimal parameters $\theta_\mr{opt}$, we compute the cost functions $C_\mr{LHST}(U^{(L)}, V^{(L)}(\theta_\mr{opt}))$ and $C_\mr{LHST}(U^{(L)}, V^{(L)}(\theta_\mr{opt}))$ to evaluate how well the ansatz $V^{(L)}(\theta_\mr{opt})$ reproduces $U^{(L)}$. 
We deal with the size $L=40$, the time $\tau = 0.5$, and the ansatz depth $d$ up to $5$. 

First, we show the numerical results for the depth compression in the intermediate size $\tilde{L}$. The compilation size $\tilde{L} = 2 \lceil l_0 + d_H + v \tau + d^\prime +1/ 2 \rceil$ should be at-least larger than $d_H +v \tau + 2d$ with considering $d=L^\prime/4 + d^\prime$.
The AFM Heisenberg Hamiltonian $H_\mr{AFM}^{(L)}$ has the range of interactions, $d_H=1$, and now we are assuming $d=5$. Since $v \tau$ is expected to be not so large under $\tau=0.5$, we choose $\tilde{L}=20$. 
We compute the cost function $C_\mr{LHST}^{(\tilde{L}/2)}(U^{(\tilde{L})},V^{(\tilde{L})})$ based on Eq. (\ref{Eq:Cost_LHST_j}) with a $2\tilde{L}$-qubit MPS having the bond dimension $30$. For implementing $U^{(\tilde{L})}$, we employ the Trotter decomposition with sufficiently large depth $d=100$,
\begin{equation}
    U^{(\tilde{L})} \simeq U_{\mr{trot},d}^{(\tilde{L})} \equiv \left( e^{-i H_\mr{even}^{(\tilde{L})} \tau/d} e^{-i H_\mr{odd}^{(\tilde{L})} \tau/d} \right)^d,
\end{equation}
where $H_\mr{odd}$ [$H_\mr{even}$] represents terms composed of interactions between $(2k-1)$-th and $2k$-th sites [$2k$-th and $(2k+1)$-th sites] in $H_\mr{AFM}$. We variationally minimize the cost function by the Broyden-Fletcher–Goldfarb–Shanno (BFGS) method implemented in SciPy \cite{2020SciPy-NMeth} with maximum iteration set to $128$. 
The initial parameter set $\theta$ is chosen as $\theta_\mr{trot}^d$ so that the ansatz $V(\theta_\mr{trot}^d)$ becomes equivalent to the Trotter decomposition with the same depth, $U_{\mr{trot},d}$, except for the global phase.

Figure \ref{fig:numerical} (a) shows the history of the cost function during the optimization in $\tilde{L}=20$, represented by the yellow ($d=3$), blue ($d=4$), red ($d=5$) solid lines in the panel. 
For the comparison, we also compute the cost functions for shallow-depth Trotter decomposition $C_\mr{LHST}^{(\tilde{L}/2)}(U^{(\tilde{L})}, V^{(\tilde{L})}(\theta_\mr{trot}^d))$ with various $d$, as described by the dashed lines.
For each depth $d=3,4,5$, the ansatz with the resulting optimal parameter set $\theta_\mr{opt}$ overwhelms the same-depth Trotter decomposition.
For instance, the $5$-depth ansatz $V^{(\tilde{L})}(\theta_\mr{opt})$ provides the cost value $7.80 \times 10^{-5}$, which is as large as that for the $40$-depth Trotter decomposition, $8.48 \times 10^{-5}$. In other words, we successfully compress the time evolution operator from depth $40$ to depth $5$ under the compilation size $\tilde{L}=20$.

Next, we examine how the larger-scale time evolution operator $U^{(L)}$ is approximated by our protocol.
Hereafter, we concentrate on the $5$-depth ansatz, and employ the corresponding optimal parameter set as $\theta_\mr{opt}$.
Considering the approximate translation invariance, the size-extended ansatz $V^{(L)}(\theta_\mr{opt})$ is constructed by copying the two-qubit gate of $V^{(\tilde{L})}(\theta_\mr{opt})$ in the spatial directions. 
We again approximate $U^{(L)}$ by the large-depth Trotter decomposition $U^{(L)}_{\mr{trot},d=100}$, and compute the cost functions as described in Fig. \ref{fig:numerical} (b). 
As Theorem \ref{Thm:local_compilation_PBC} says, the local cost functions $C_\mr{LHST}$ and $C_\mr{LHST}^{(L/2)}$ (the purple and brown solid lines) hardly increase when we employ $\theta_\mr{opt}$ in $\tilde{L}=20$ for the larger-scale ansatz with $L \geq 20$. 
Reflecting the fact that Theorem \ref{Thm:local_compilation_PBC} yields the loose bound proportional to $L$, the global cost function $C_\mr{HST}$ (the red solid line) experiences a gradual increase in $L$, but remains sufficiently small compared to $1$.
Any cost function for the ansatz with $\theta_\mr{opt}$ is comparably smaller than that with $\theta_\mr{trot}^{d=5}$, the parameter set for reproducing the Trotter decomposition with the same depth $d=5$ (See the blue, orange, and light-green solid lines).

We also assess the average gate fidelity. Based on Eqs. (\ref{Eq:operational_meaning_HST}) and (\ref{Eq:operational_meaning_LHST}), the ansatz extended to $L=40$ qubits is ensured to have $\bar{F}(U^{(L)},V^{(L)}(\theta_\mr{opt})) \geq 0.9977$, while the same-depth Trotter decomposition provides $\bar{F}(U^{(L)},V^{(L)}(\theta_\mr{trot}^{d=5})) \geq 0.8580$.
Therefore, our protocol succeeds in implementing the time evolution operator for the larger-scale $20 \leq L \leq 40$ with the limited depth by exploiting the local compilation on the size $\tilde{L}=20$.

\begin{figure}[t]
    \includegraphics[height=6.43cm, width=8.5cm]{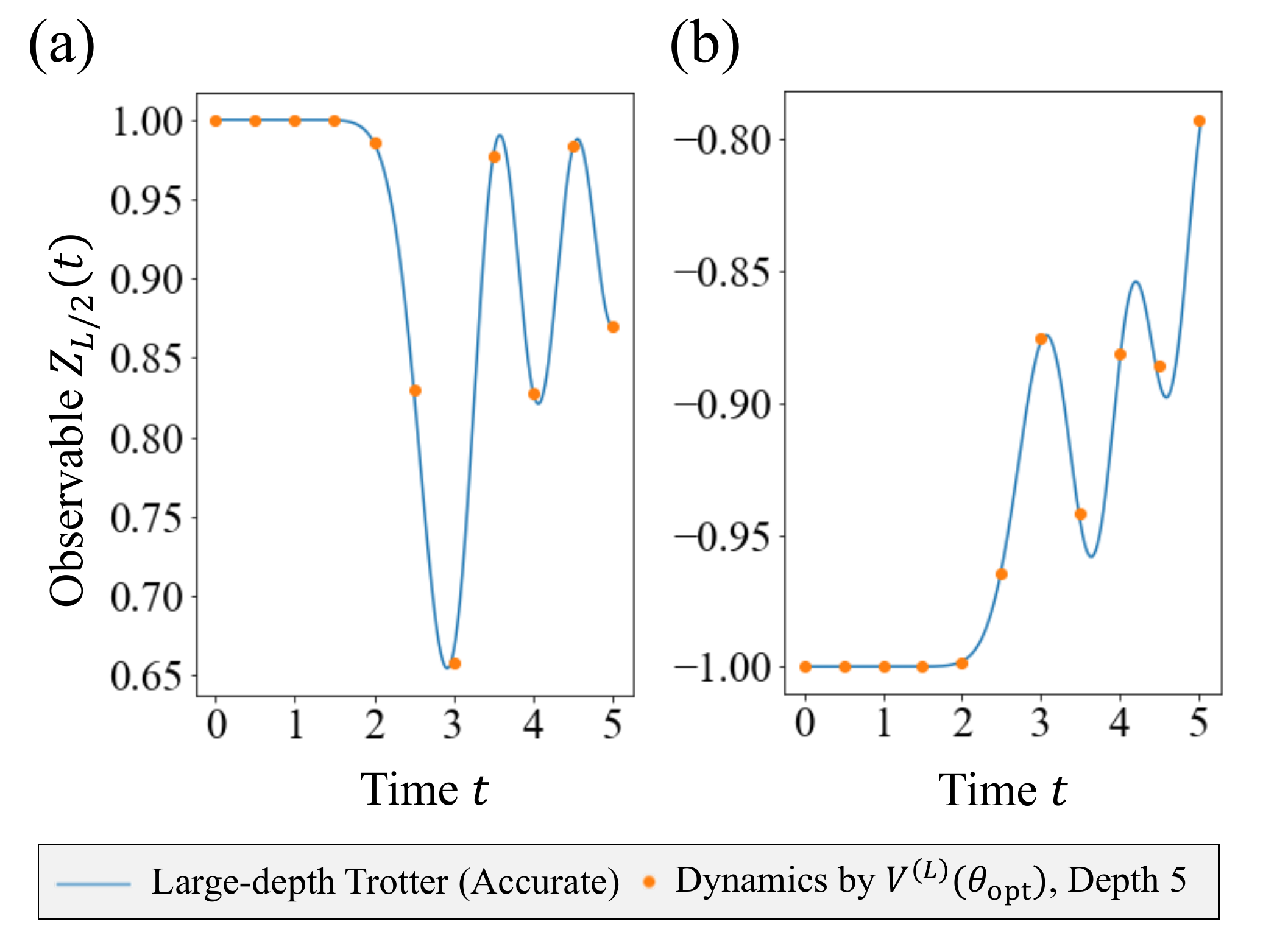}
    \caption{Real-time dynamics of $Z_{L/2}$ (a) from the ferromagnetic initial state with two local excitations $\ket{\psi_\mr{LE}^{(L)}(0)}$ and (b) from the one with two domain walls $\ket{\psi_\mr{DW}^{(L)}(0)}$. The orange dots represents the stroboscopic dynamics at $t \in \tau \mathbb{Z}$ under $V^{(L)}(\theta_\mr{opt})$, implemented with the depth $50$ up to $t=5$. This well corresponds to the blue line, which shows an accurate dynamics under the Trotter decomposition with sufficiently large depth $100$ per $\tau=0.5$.} 
    \label{fig:numerical_dynamics}
\end{figure}

Finally, we demonstrate how well the compressed time evolution operator $V^{(L)}(\theta_\mr{opt})$ reproduces the dynamics of larger-scale systems under the accurate one $U^{(L)}$. 
By applying $V^{(L)}(\theta_\mr{opt})$ or its inverse repeatedly, we can approximately simulate the stroboscopic dynamics at the time $t \in \tau \mathbb{Z}$, which is larger than the original time scale $\tau$, with a smaller-depth circuit. 
Furthermore, it should be noted that our protocol can capture larger-scale phenomena in the size $L$ despite the compilation in $\tilde{L} < L$.
To confirm this numerically, we simulate the stroboscopic dynamics which involves the time scale and the size scale respectively larger than $\tau=0.5$ and $\tilde{L}=20$.

As the simplest cases, we prepare the following two initial states,
\begin{eqnarray}
    \ket{\psi^{(L)}_\mr{LE}(0)} &=& X_{(L-\tilde{L})/2} X_{(L+\tilde{L})/2} \ket{0}^{\otimes L}, \\
    \ket{\psi^{(L)}_\mr{DW}(0)} &=& \left( \prod_{j = (L-\tilde{L})/2}^{(L+\tilde{L})/2} X_j \right) \ket{0}^{\otimes L},
\end{eqnarray}
for the size $L=40$.
They respectively represent ferromagnetic states having two local excitations (for $\ket{\psi_\mr{LE}(0)}$) and two domain walls (for $\ket{\psi_\mr{DW}(0)}$) with distance $\tilde{L}=20$. 
Then, we evaluate the expectation value of $Z_{L/2}$ evolving under the Hamiltonian $H_\mr{AFM}^{(L)}$. 
Intuitively, it is expected that two distant local excitations or domain walls at the $(L-\tilde{L})/2$-th and $(L+\tilde{L})/2$-th sites respectively propagate in both left and right directions under the Hamiltonian $H_\mr{AFM}^{(L)}$, and the central site $j=L/2$ observes their collisions. Thus, the change in the expected value of $Z_{L/2}$ can be employed as a diagnosis for the larger-scale dynamics involving at-least $\tilde{L}+1$ sites, which is larger than the compilation size.

Figure \ref{fig:numerical_dynamics} shows the numerical results for the approximate stroboscopic dynamics obtained by the compilation.  
With the $5$-depth ansatz $V^{(L)}(\theta_\mr{opt})$ obtained by the optimization in the size  $\tilde{L}=20$, we compute the state and its local observable, given by
\begin{eqnarray}
    \ket{\psi^{(L)}(n \tau)} &=& V^{(L)}(\theta_\mr{opt})^n \ket{\psi^{(L)}(0)}, \quad n \in \mathbb{N}, \\
    Z_{L/2}(n \tau) &=& \braket{\psi^{(L)}(n \tau) | Z_{L/2} | \psi^{(L)}(n \tau)}.
\end{eqnarray}
We employ MPS with the bond dimension $60$ for simulating the dynamics from the initial states $\ket{\psi_\mr{LE}(0)}$ or $\ket{\psi_\mr{DW}(0)}$, which are depicted as the orange dots respectively in Fig. \ref{fig:numerical_dynamics} (a) and (b).
We also compute the dynamics under the large-depth Trotter decomposition $U^{(L)}_{\mr{trot},d=100}$ as the accurate dynamics for the comparison (See the blue solid lines).
In both cases, the compilation results well reproduce the accurate dynamics up to $t \leq 10 \tau = 5$ with the mean square errors $5.27 \times 10^{-6}$ and $1.29 \times 10^{-6}$ \cite{remark:numerical_dynamics}.
We conclude that our prescription exploiting the intermediate-size $\tilde{L}$ and the fixed time $\tau$ provides an appropriate shallow-depth time evolution operator useful for larger-scale quantum systems both in space and time.
We also remark that the optimal parameter obtained here is expected to be useful for even larger-scale quantum simulations beyond the size considered in this work from the size-dependence in Fig. \ref{fig:numerical} (b).
Our numerical results suggest the feasibility of the classical local compilation to design large-scale quantum circuits, in addition to the possible quantum local compilation by NISQ devices.
 
\section{Discussion and Conclusion}\label{Sec:Discussion}

In this paper, we develop the local variational quantum compilation (LVQC), in which we conduct local optimization for intermediate-scale quantum systems designated by the Lieb-Robinson bound, and obtain an approximate time evolution operator of larger-scale quantum systems. 
Since the approximation error of the local cost function supporting our protocol relies only on the Lieb-Robinson bound, it has broad applicability to finite-ranged, short-ranged, and long-ranged interacting large-scale systems in generic dimension.
LVQC begins with the local compilation by intermediate-scale quantum devices or corresponding classical simulators, and ends up with the quantum execution of the compiled larger-scale dynamics.
Therefore, not only it unveils a classical approach to design large-scale quantum circuits, but also it will play a significant role in bridging NISQ device technique to the practical use of larger quantum devices as the long-term goal.

We finish this article with providing some future directions.
The first one is to seek for the possibility of the local compilation in classical ways. 
While we refer to our protocol as ``quantum" compilation, Theorems \ref{Thm:local_compilation_PBC} and \ref{Thm:Local_compilation_generic} are not limited to the context of variational quantum algorithms where we optimize parametrized quantum circuit in a quantum-classical hybrid manner.
Our numerical demonstration based on TEBD involving up to $40$ qubits is indeed a good example for using classical simulator for LVQC.
Other sophisticated techniques (e.g. tensor-network-based methods for 2D systems) will also be important for executing our protocol on classical computers.
We might also be able to use exact brute-force classical simulators for LVQC in future.
This is because, for finite-ranged or short-ranged systems with $v\tau = O(L^0)$, LVQC ensures the classical efficient evaluaion of the cost function in time $e^{O(\tilde{L})} = \mr{poly} (L)$ given that it is sufficient to take $\tilde{L} = O(\log L)$ in this case.
Although current classical devices are still not capable of simulating quantum systems with size $\tilde{L}$, which typically becomes more than tens of qubits, LVQC without resorting to approximate simulators may be available in future.
Note that this does not contradict with the existing result that states the evaluation of the cost functions $C_\mr{LHST}$ and $C_\mr{HST}$ in polynomial accuracy with respect to the system size $L$ for general unitaries is a DQC1-hard problem \cite{Khatri2019-gj} (DQC1; efficiently solvable problems by one clean qubit and other noisy qubits \cite{Knill1998-xs}), since we restrict ourselves to certain short-time local Hamiltonian dynamics and shallow depth ansatzes  (See Appendix \ref{Asec:dqc1} for detail).

The second significant task for future is to accumulate benchmark results by both classical and quantum simulation, including higher-dimensional cases, short-ranged interacting cases, and long-ranged interacting cases.
Several programmable quantum simulators, such as superconducting qubits \cite{chow2021ibm} and Rydberg atoms \cite{Ebadi2022-uj}, have recently achieved a few hundred qubits with high-controllability and two-dimensionality, and they will be available for both the local compilation and the quantum execution of the compressed time evolution. 
For instance, as an immediate task to be tackled, it may be possible to observe long-time dynamics beyond the current coherence time on such compiled quantum simulators by the classical local compilation for tens of qubits.
One of the ultimate goals is to compile time evolution operators for huge quantum chemistry materials such as molecules and crystals. 
Although long-ranged Hamiltonians of electrons from the first principles are out of scope with the current knowledge of the LR bound, our protocol is expected to be valid for various materials under the reorganization of approximate models based on their structures.
As for including the improvement for long-ranged cases, we leave it as future work.

\section*{Acknowledgment}
Kaoru Mizuta is supported by WISE Program, MEXT, and a Research Fellowship for Young Scientists from JSPS (Grants No. JP20J12930).
Kosuke Mitarai is supported by JST PRESTO Grant No. JPMJPR2019 and JSPS KAKENHI Grant No. 20K22330.
K.F. is supported by JST ERATO Grant No. JPMJER1601 and JST CREST Grant No. JPMJCR1673.
This work is supported by MEXT Quantum Leap Flagship Program (MEXTQLEAP) Grant No. JPMXS0118067394 and JPMXS0120319794.
We also acknowledge support from JST COI-NEXT program Grant No. JPMJPF2014.
A part of this work was performed for Council for Science, Technology and Innovation (CSTI), Cross-ministerial Strategic Innovation Promotion Program (SIP), “Photonics and Quantum Technology for Society 5.0” (Funding agency: QST).

\bibliography{bibliography.bib}

\begin{thebibliography}{61}
\bibitem [{rem()}]{remark:numerical_dynamics}%
  \BibitemOpen
  \href@noop {} {}\bibinfo {note} {We also compute the stroboscopic dynamics
  under the same-depth Trotterization $V^{(L)}(\theta_\mathrm{trot}^{d=5})$. As
  the mean square errors during $0 \leq t \leq 10 \tau$, we obtain $3.72 \times
  10^{-5}$ for the initial state $\ket{\psi_\mathrm{LE}^{(L)}(0)}$ and $ 7.59
  \times 10^{-6}$ for $\ket{\psi_\mathrm{DW}^{(L)}(0)}$. They are accidentally
  comparable to the LVQC results by $V^{(L)}(\theta_\mr{opt})$, in spite of the
  relatively small average gate fidelity for
  $V^{(L)}(\theta_\mathrm{trot}^{d=5})$. This comes from the choice of initial
  states, since both the accurate time evolution operator and the approximate
  one by Trotterization trivially act on the ferromagnetic regions with
  suppressing the errors.}\BibitemShut {Stop}%
\end{thebibliography}

\appendix
\begin{center}
\bf{\large Appendix}
\end{center}

\section{Proof of Lemma \ref{Thm:LqubitLHST}}\label{Asec:LqubitLHST}

The quantity we wish to evaluate is 
\begin{equation}
    \braket{\Pi_j} = \bra{\Phi_+}_{AB}(U_A \otimes V_B^{*})^\dagger \Pi_j (U_A \otimes V_B^{*})\ket{\Phi_+}_{AB}.
\end{equation}
Noting that $(U_A \otimes V_B^{*})\ket{\Phi_+}_{AB} = (U_A V^\dagger_A \otimes I_B)\ket{\Phi_+}_{AB}$ and $\Pi_j$ can be decomposed as a sum of Pauli operator by Eq. (\ref{Eq:Pi_j_decompose}), it is sufficient to evaluate
\begin{equation}
    \bra{\Phi_+}_{AB}(V_A U_A^\dagger \otimes I_B) (O_{A_j} \otimes O_{B_j}) (U_A V_A^\dagger \otimes I_B)\ket{\Phi_+}_{AB},
\end{equation}
for $O=X,Y,Z$ to obtain $\braket{\Pi_j}$.
Therefore, if we have efficient means to evaluate
\begin{equation}\label{Aeq:Bell-pair-pauli-exp-value}
    \begin{split}
    &F(W_A, P_A, P_B) \\
    &:= \bra{\Phi_+}_{AB}(W_A^\dagger \otimes I_B) P_A \otimes P_B (W_A \otimes I_B)\ket{\Phi_+}_{AB}
    \end{split}
\end{equation}
for arbitrary $L$-qubit unitary $W_A$ and Pauli operator $P_A$ and $P_B$, we can obtain $\braket{\Pi_j}$.
Here, we provide an efficient algorithm to estimate Eq. (\ref{Aeq:Bell-pair-pauli-exp-value}).

First, we observe the following equality holds:
\begin{align}
    &F(W_A, P_A, P_B) \nonumber\\
    &= \frac{1}{2^L}\sum_{i,j=1}^{2^L} \bra{i}_AW_A^\dagger P_A W_A \ket{j}_A \bra{i}_BP_B\ket{j}_B \\
    &= \frac{1}{2^L}\sum_{i,j=1}^{2^L} \mathrm{Re}\left[\bra{i}_AW_A^\dagger P_A W_A \ket{j}_A \bra{i}_BP_B\ket{j}_B\right]\label{Aeq:decompose-LHST}
\end{align}
where we used the definition of the Bell state $\ket{\Phi_+}_{AB}=\sum_{i=1}^{2^L}\ket{i}_A\ket{i}_B/\sqrt{2^L}$ and that $F(W_A, P_A, P_B)$ is real.
Now, a Monte-Carlo approach can be employed to evaluate the sum of Eq. (\ref{Aeq:decompose-LHST}).

The algorithm we propose is as follows. 
First, sample $x$ from the uniform distribution on $\{1,2,...,2^L\}$.
Let $\alpha_x\ket{y_x}= P_B\ket{x}$ where $\ket{y_x}$ and $\alpha_x\in\{\pm 1, \pm i\}$ is a computational basis and a coefficient determined by $x$ and $P_B$.
Then, we estimate $\bra{y_x}W_A^\dagger P_A W_A\ket{x}$ on an $L$-qubit quantum device within an addtive error $\epsilon$.
This can be achieved by utilizing the following equalities that holds for an arbitrary observable $O$:
\begin{align}
    2\mathrm{Re}[\bra{y}O\ket{x}] &= \bra{+_{x,y}}O\ket{+_{x,y}} - \bra{-_{x,y}}O\ket{-_{x,y}},\label{Aeq:matrix-element-re}\\
    2\mathrm{Im}[\bra{y}O\ket{x}] &= \bra{+i_{x,y}}O\ket{+i_{x,y}} - \bra{-i_{x,y}}O\ket{-i_{x,y}},\label{Aeq:matrix-element-im}
\end{align}
where $\ket{\pm_{x,y}}:=(\ket{x}\pm\ket{y})/\sqrt{2}$ and $\ket{\pm i_{x,y}}:=(\ket{x}\pm i\ket{y})/\sqrt{2}$.
More precisely, for a given pair $(x,y_x)$, we first evaluate expectation values $\bra{\pm_{x,y_x}}W_A^\dagger P_A W_A\ket{\pm_{x,y_x}}$ or $\bra{\pm i_{x,y_x}}W_A^\dagger P_A W_A\ket{\pm i_{x,y_x}}$ using $N_1$ samples each and then combine them according to the above formula.
Let an estimator of $\bra{y_x}W_A^\dagger P_A W_A\ket{x}$ obtained by this procedure be $\hat{P}_{A,x}$.
Importantly, $\mathrm{Var}[\hat{P}_{A,x}]=\mathcal{O}(1/N_1)$.
Finally, we construct an estimator of $F(W_A, P_A, P_B)$ as
\begin{align}
    \hat{F}(W_A, P_A, P_B) := \mathrm{Re}\left[\alpha_x \hat{P}_{A,x}\right].
\end{align}
From this form of the estimator, it is sufficient to evaluate only $\mathrm{Re}[\bra{y}O\ket{x}]$ ($\mathrm{Im}[\bra{y}O\ket{x}]$) by Eqs. (\ref{Aeq:matrix-element-re}) and (\ref{Aeq:matrix-element-im})  when $\alpha_x$ is real (imaginary).
Note that $\hat{F}(W_A, P_A, P_B)$ is defined by two random variables $x$ and $\hat{P}_{A,x}$.

To see that $\hat{F}(W_A, P_A, P_B)$ is indeed an efficient unbiased estimator, we analyze its expectation value and variance.
Let us assume that, for a fixed $x$, the random variable $\mathrm{Re}[\alpha_x\hat{P}_{A,x}]$ follows a probability distribution $p_x(a)$.
The probability that $\hat{F}(W_A, P_A, P_B)$ takes a specific value $f$ is given by $\sum_{x}p_x(f)/2^L$.
Then, we can calculate $\mathbb{E}[\hat{F}(W_A, P_A, P_B)]$ and $\mathbb{E}[\hat{F}(W_A, P_A, P_B)^2]$ as follows:
\begin{align}
    &\mathbb{E}[\hat{F}(W_A, P_A, P_B)] \nonumber\\
    &= \sum_{x}\sum_{f} f \frac{p_x(f)}{2^L} \nonumber\\
    &= \frac{1}{2^L}\sum_x \mathbb{E}_{a\sim p_x(a)}[a] \nonumber\\
    &= \frac{1}{2^L}\sum_x \mathrm{Re}\left[\bra{y_x}_AW_A^\dagger P_A W_A \ket{x}_A \bra{y_x}_BP_B\ket{x}_B\right] \nonumber\\
    &= \frac{1}{2^L}\sum_{x,y} \mathrm{Re}\left[\bra{y}_AW_A^\dagger P_A W_A \ket{x}_A \bra{y}_BP_B\ket{x}_B\right] \nonumber\\
    &= F(W_A, P_A, P_B),\label{Aeq:unbiasedF}
\end{align}
\begin{align}
    &\mathbb{E}[\hat{F}(W_A, P_A, P_B)^2] \nonumber \\
    &= \sum_{x}\sum_{f}  f^2 \frac{p_x(f)}{2^L} \nonumber\\
    &= \frac{1}{2^L}\sum_{x}\left[\mathrm{Var}_{a\sim p_x}[a]+\sum_x \mathbb{E}_{a\sim p_x}[a]^2\right] \nonumber\\
    &\leq \max_x \mathrm{Var}_{a\sim p_x}[a^2] \\
    &\quad + \frac{1}{2^L}\sum_x \mathrm{Re}\left[\bra{y_x}_AW_A^\dagger P_A W_A \ket{x}_A \bra{y_x}_BP_B\ket{x}_B\right]^2
\end{align}
Equation (\ref{Aeq:unbiasedF}) shows that $\hat{F}(W_A, P_A, P_B)$ is an unbiased estimator of $F(W_A, P_A, P_B)$, the desired quantity.
Combining the above with 
\begin{equation}
    \mathrm{Var}_{a\sim p_x(a)}[a^2]=\mathbb{E}[a^2]-\bra{y_x}W_A^\dagger P_A W_A\ket{x}^2 = \mathcal{O}(1/N_1)
\end{equation}
for all $x$, we obtain,
\begin{align}
    \mathrm{Var}[\hat{F}(W_A, P_A, P_B)] \leq \mathcal{O}(1/N_1)+\mathcal{V},
\end{align}
where
\begin{align}
\begin{split}
    \mathcal{V}&:= \sum_x \mathrm{Re}\left[\bra{y_x}_AW_A^\dagger P_A W_A \ket{x}_A^2 \bra{y_x}_BP_B\ket{x}_B\right]^2 \\
    &\quad- \left(\sum_x \mathrm{Re}\left[\bra{y_x}_AW_A^\dagger P_A W_A \ket{x}_A \bra{y_x}_BP_B\ket{x}_B\right]\right)^2,
\end{split}
\end{align}
is the variance of this protocol when we can exactly estimate $\mathrm{Re}[\bra{y_x}_AW_A^\dagger P_A W_A \ket{x}_A \bra{y_x}_BP_B\ket{x}_B]$.

Since $\mathrm{Re}[\bra{y_x}_AW_A^\dagger P_A W_A \ket{x}_A \bra{y_x}_BP_B\ket{x}_B]=\mathcal{O}(1)$, $\mathcal{V}$ is also $\mathcal{O}(1)$.
This implies that a sample mean of $N_2$ independent samples of $\hat{F}(W_A,P_A,P_B)$, which requires $N=N_1N_2$ runs of quantum devices for its construction, has variance $\mathcal{O}(1/(N_1N_2))+\mathcal{O}(1/N_2)$.
Therefore, it is sufficient to take $N_1=\mathcal{O}(1)$, $N_2=\mathcal{O}(1/\epsilon^2)$ and thus $N=\mathcal{O}(1/\epsilon^2)$ to obtain an estimate of $F(W_A,P_A,P_B)$ within an additive error $\epsilon$ with high probability.

The same strategy can be taken to evaluate $C_{\mathrm{HST}}(U,V)$.
In this case, the task is to estimate the expectation value of $\Pi_1\Pi_2\cdots\Pi_L$ with respect to $(U_A\otimes V_B^*)\ket{\Phi_+}_{AB}$.
We use the fact that $\Pi_1\Pi_2\cdots\Pi_L$ can also be expanded as a sum of Pauli operator:
\begin{align}
    \Pi_1\Pi_2\cdots\Pi_L = \frac{1}{4^L}\sum_{P\in\{I,X,Y,Z\}^{\otimes L}} c_P P\otimes P,
\end{align}
where $c_P=1$ when $P$ has even number of $Y$ and $c_P=-1$ otherwise.
This decomposition has exponential number of Pauli operators, and a naive approach where we estimate expectation values of every Pauli operator takes exponential time to $L$.
However, we can take an Monte-Carlo approach to evaluate this sum by interpreting the coeffcient $1/4^L$ as a proability.

The algorithm for evaluating $C_{\mathrm{HST}}(U,V)$ is as follows. 
First, we pick up a Pauli operator $P\in\{I,X,Y,Z\}^{\otimes L}$ randomly.
Then, we estimate the expectation value of $P\otimes P$ using the algorithm in the proof of Lemma \ref{Thm:LqubitLHST}.
Repeating the above procedure $N_3=\mathcal{O}(1/\epsilon^2)$ times while setting $N_1,N_2=\mathcal{O}(1)$, we obtain $C_{\mathrm{HST}}(U,V)$ within an additive error $\epsilon$ with high probability using $N=N_1N_2N_3=\mathcal{O}(1/\epsilon^2)$ samples in total. \hfill $\square$

\section{Extension to other cases}\label{Asec:Extension}
In the main text, we mainly focus on one-dimensional systems with finite-ranged interactions. Here, we discuss the extensions of our results to other cases in terms of the range of interactions and the dimension of systems.

From the derivation of Theorems \ref{Thm:local_compilation_PBC} and \ref{Thm:Local_compilation_generic} in the main text, the range of interactions and the dimension affect our results only via $\varepsilon_\mr{LR}$ in Eq. (\ref{Eq:Def_epsilon_LR1}), coming from LR bound. To be precise, we should change the choice of the intermediate size $L^\prime = 2 (l_0 + d_H + v \tau)$ or $\tilde{L} \geq L^\prime + 2d^\prime + 1$, which designates the restriction of the Hamiltonian and the ansatz, so that the bound $\varepsilon_\mr{LR}$ can be ignored. Thus, after deriving $\varepsilon_\mr{LR}$ caused by the Hamiltonian restriction in Appendix \ref{Asubsec:Hamiltonian_restriction}, we devote the following sections \ref{Asubsec:Extension_Finite_range}-\ref{Asubsec:Extension_Long_range} to discuss an appropriate choice of the size for finite-ranged, short-ranged, and long-ranged cases in generic dimension.

\subsection{Hamiltonian restriction by Lieb-Robinson bound}\label{Asubsec:Hamiltonian_restriction}
We first discuss the error bound $\varepsilon_\mr{LR}$ in Eq. (\ref{Eq:Def_epsilon_LR1}), caused by the restriction of Hamiltonian to a local terms around a site $j$. Let us assume that a Hamiltonian $H$ has the LR bound designated by
\begin{equation}\label{Aeq:LiebRobinsonBound}
    \| [e^{i H \tau} O_X e^{-i H \tau}, O_Y ] \| \leq \|O_X\| \cdot \|O_Y \| \cdot \mathcal{C}(\mr{dist}(X,Y),\tau),
\end{equation}
for local observables $O_X$ and $O_Y$, whose supports are respectively the subsets of the lattice, $X$ and $Y$ ($\subseteq \Lambda$). The distance between domains is defined by
\begin{equation}
    \mr{dist}(X,Y) = \inf \{ \mr{dist}(j,j^\prime) \, | \, j \in X, \, j^\prime \in Y \}.
\end{equation}
We also define the distance between a site $j$ and a domain $Y$ by $\mr{dist}(j,Y)=\mr{dist}(X=\{j\},Y)$.

Assuming the existence of the LR bound, we consider the dynamics of local observables.
We define the restriction of the Hamiltonian $H^{(L)}= \sum_X h_X$ for generic $D$-dimensional systems by
\begin{eqnarray}
    H^{(L^\prime,j)} &=& \sum_{X; X \subseteq \Lambda_{L^\prime,j}} h_X, \label{Aeq:Def_Restricted_Hamiltonian} \\
    \Lambda_{L^\prime,j} &=& \{ j^\prime \in \Lambda \, | \, \mr{dist}(j,j^\prime) \leq L^\prime /2 \}, \label{Aeq:Def_Restricted_domain}
\end{eqnarray}
where $L$ and $L^\prime$ ($\leq L$) respectively represent the linear scales of the lattices $\Lambda$ and $\Lambda_{L^\prime,j}$. 
It is expected that the dynamics of local observables, $e^{i H^{(L)} \tau} O_j e^{-iH^{(L)}\tau}$, is well described by the restricted Hamiltonian $H^{(L^\prime,j)}$ for sufficiently large $L^\prime$, and in fact, it has been proved by Refs. \cite{Robinson1976-wd,Nachtergaele2006-ok,Nachtergaele2006-il} for finite-ranged and short-ranged cases. 
In order to cover long-ranged cases and make our paper self-contained, we summarize and rederive the result in a slightly different way below. 
After that, we derive proper choice of the compilation size $\tilde{L}$ for finite-ranged, short-ranged, and long-ranged cases in generic dimension.

\begin{lemma}\label{Alemma:Hamiltonian_restriction}
We assume the existence of the LR bound in the form of Eq. (\ref{Aeq:LiebRobinsonBound}) on the Hamiltonian $H^{(L)}$, and define the size of a domain $X \subseteq \Lambda$ by
\begin{equation}
    r(X) = \max \{ \mr{dist}(j,j^\prime) \, | \, j, j^\prime \in X \}.
\end{equation}
When the function $\mathcal{C}(r,t)$ is monotonically decreasing in the distance $r$ and monotonically increasing in the time $\tau$, the inequality
\begin{equation}\label{Aeq:Def_epsilon_LR}
    \| e^{i H^{(L)} \tau} O_j e^{-i H^{(L)} \tau} - e^{i H^{(L^\prime,j)} \tau} O_j e^{-i H^{(L^\prime,j)} \tau} \| \leq \varepsilon_\mr{LR},
\end{equation}
\begin{eqnarray}
    \varepsilon_\mr{LR} &=& C_1 \int_{L^\prime/2 - r_H}^\infty r^{D-1} \mathcal{C}(r,\tau) dr + \varepsilon (r_H), \label{Aeq:Result_epsilon_LR} \\
    \varepsilon(r_H) &=& C_2 \sum_{i \in \Lambda_{L^\prime,j}} \, \sum_{X; X \ni i, r(X) >  r_H} \| h_X \| \label{Aeq:Result_epsilon_rH}
\end{eqnarray}
is satisfied, where the length scale $r_H$ is an arbitrary value satisfying $0 \leq r_H \leq L^\prime /2$, and the constants $C_1$ and $C_2$ are independent of $L$ and $L^\prime$.
\end{lemma}

\textit{Proof.}--- The proof is mainly based on Ref. \cite{Else2020-tr}, but we make a slight change so that it can cover short-ranged and long-ranged interactions. First, we define a function $f(t)$ by
\begin{equation}
    f(t) = O_j - U^{(L^\prime,j)}_t U^{(L) \dagger}_t O_j U^{(L)}_t U^{(L^\prime,j) \dagger}_t,
\end{equation}
\begin{equation}
    U^{(L)}_t = e^{-i H^{(L)}t}, \quad U^{(L^\prime,j)}_t = e^{ - i H^{(L^\prime,j)} t}.
\end{equation}
$\|f(\tau)\|$ equals the left hand side of Eq. (\ref{Aeq:Def_epsilon_LR}). Then, the differentiation of $f(t)$ in $t$ immediately results in
\begin{equation}
    f^\prime (t) = i U^{(L^\prime,j)}_t \left[ U^{(L)\dagger}_t O_j U^{(L)}_t, H^{(L)}-H^{(L^\prime,j)} \right] U^{(L^\prime,j)\dagger}_t.
\end{equation}
Considering that $f(0)=0$, the operator norm $\|f(\tau)\|$ is bounded from above as follows;
\begin{eqnarray}
    \| f(\tau) \| &=& \left\| \int_0^\tau f^\prime(t) dt \right\| \leq
    \int_0^\tau  \left\| f^\prime(t) \right\| dt \nonumber \\
    &=& \int_0^\tau  \left\| \left[ U^{(L)\dagger}_t O_j U^{(L)}_t, H^{(L)}-H^{(L^\prime,j)} \right] \right\| dt. \nonumber \\
    && \label{Aeq:Norm_f_tau}
\end{eqnarray}

From the definition of $H^{(L^\prime,j)}$, given by Eq. (\ref{Aeq:Def_Restricted_Hamiltonian}), we obtain
\begin{equation}
    H^{(L)}-H^{(L^\prime,j)} = \sum_{X; X \nsubseteq \Lambda_{L^\prime,j}} h_X.
\end{equation}
Introducing an arbitrary length scale $r_H$, satisfying $0 \leq r_H \leq L^\prime/2$, the summation over $X$, which is not a subset of $\Lambda_{L^\prime,j}$, can be divided in the following way,
\begin{equation}
    \sum_{X; X \nsubseteq \Lambda_{L^\prime,j}} = \sum_{X \in \mathcal{X}_A} + \sum_{X \in \mathcal{X}_B(r_H)} + \sum_{X \in \mathcal{X}_C(r_H)},
\end{equation}
where each of $\mathcal{X}_A$, $\mathcal{X}_B(r_H)$, and $\mathcal{X}_C(r_H)$, is defined by
\begin{eqnarray}
    \mathcal{X}_A &=& \{ X \, | \, X \subseteq \Lambda \backslash \Lambda_{L^\prime,j} \}, \label{Aeq:Domain_X_A} \\
    \mathcal{X}_B(r_H) &=& \{ X \nsubseteq \Lambda_{L^\prime,j} \, | \, X \cap \Lambda_{L^\prime,j} \neq \phi, \, r(X) \leq r_H  \}, \nonumber \\
    && \label{Aeq:Domain_X_B} \\
    \mathcal{X}_C(r_H) &=& \{ X \nsubseteq \Lambda_{L^\prime,j} \, | \, X \cap \Lambda_{L^\prime,j} \neq \phi, \, r(X) > r_H  \}. \nonumber \\
    && \label{Aeq:Domain_X_C}
\end{eqnarray}
Using the triangular inequality of the operator norm, Eq. (\ref{Aeq:Norm_f_tau}) is further bounded by
\begin{eqnarray}
\| f(\tau) \| &\leq& \varepsilon_{AB} (r_H) + \varepsilon_C(r_H), \\
\varepsilon_{AB} (r_H) &=& \sum_{X \in \mathcal{X}_A \cup \mathcal{X}_B (r_H)} \int_0^\tau  \left\| \left[ U^{(L)\dagger}_t O_j U^{(L)}_t, h_X \right] \right\| dt, \nonumber\\
&& \\
\varepsilon_C(r_H) &=& \sum_{X \in \mathcal{X}_C (r_H)} \int_0^\tau  \left\| \left[ U^{(L)\dagger}_t O_j U^{(L)}_t, h_X \right] \right\| dt. \nonumber \\
&&
\end{eqnarray}
We now evaluate the upper bound of $\varepsilon_{AB} (r_H)$ and that of $\varepsilon_C(r_H)$, respectively.  

For the first term $\varepsilon_{AB} (r_H)$, we use the fact that a domain $X$, which belongs to $\mathcal{X}_A \cup \mathcal{X}_B (r_H)$, satisfies $\mr{dist}(j,X) \geq L^\prime/2 - r_H$ from their constructions Eqs. (\ref{Aeq:Domain_X_A}) and (\ref{Aeq:Domain_X_B}). Using the LR bound Eq. (\ref{Aeq:LiebRobinsonBound}) for the integrand, $\varepsilon_{AB} (r_H)$ is bounded by
\begin{eqnarray}
 && \sum_{X \in \mathcal{X}_A \cup \mathcal{X}_B (r_H)} \int_0^\tau \| O_j \| \cdot \| h_X \| \cdot \mathcal{C}(\mr{dist}(j,X),t) dt, \nonumber \\
 &\leq& \sum_{j^\prime; \mr{dist}(j,j^\prime)\geq L^\prime/2 - r_H} \|O_j \| \sum_{X; X \ni j^\prime} \tau \| h_X \| \cdot \mathcal{C}(\mr{dist}(j,j^\prime),\tau), \nonumber \\
 &\leq& g \tau \| O_j \| \sum_{j^\prime; \mr{dist}(j,j^\prime)\geq L^\prime/2 - r_H} \mathcal{C}(\mr{dist}(j,j^\prime),\tau). \label{Aeq:Bound_epsilon_AB}
\end{eqnarray}
In the first inequality, we employ the monotonicity of $\mathcal{C}(r,t)$, which validates the replacement by $\mathcal{C}(\mr{dist}(j,X),t) \leq \mathcal{C}(\mr{dist}(j,j^\prime),\tau)$ for $X \ni j^\prime$ and $t \leq \tau$.
For the second inequality, we use Eq. (\ref{Eq:Extensiveness}).
Concerning the summation over $j^\prime$ in the last line, the number of sites $j^\prime$ satisfying $\mr{dist}(j,j^\prime) \simeq r$ is proportional to the surface area $S_D r^{D-1}$ under the finite density $\rho$. Thus, the summation $\sum_{j^\prime; \mr{dist}(j,j^\prime)\geq L^\prime/2 - r_H}$ is expected to be approximated by $\int_{L^\prime/2-r_H}^\infty dr \rho S_D r^{D-1}$.
As a matter of fact, following this intuition, when $\mathcal{C}(r,t)$ is monotonically decreasing in $r$ and the number of sites per volume is finite, there exists a positive constant $C_3$ such that
\begin{equation}\label{Aeq:Integral_LR_bound}
    [\text{Eq. (\ref{Aeq:Bound_epsilon_AB})}] \leq g \tau \| O_j \| \cdot C_3 \int_{L^\prime/2-r_H}^\infty r^{D-1} \mathcal{C}(r,\tau) dr,
\end{equation}
for generic $D$-dimensional systems \cite{Else2020-tr}. Here, the constant $C_3$ depends only on the dimension and the density of the lattice, but not on $L$ and $L^\prime$. Defining the constant $C_1$ by $C_1=g\tau \|O_j\| C_3$, $\varepsilon_{AB}(r_H)$ is bounded from above by the first term in the right hand side of Eq. (\ref{Aeq:Result_epsilon_LR}).

For the second term $\varepsilon_C(r_H)$, we soon arrive at
\begin{eqnarray}
    \varepsilon_C(r_H) &\leq& \sum_{X \in \mathcal{X}_C (r_H)} 2 \tau \| O_j \| \cdot \| h_X \|, \nonumber \\
    &\leq& 2 \tau \| O_j \| \sum_{i \in \Lambda_{L^\prime,j}} \sum_{X; X \ni i, r(X)>r_H} \| h_X \|, \nonumber \\
    &&
\end{eqnarray}
where we use the definition of $\mathcal{X}_C(r_H)$, Eq. (\ref{Aeq:Domain_X_C}), to derive the second inequality. When we choose a constant $C_2$ by $2\tau \|O_j\|$, which is independent of $L$ and $L^\prime$, $\varepsilon_C(r_H)$ is bounded by $\varepsilon(r_H)$ [See Eq. (\ref{Aeq:Result_epsilon_rH})] from above.

Combining these upper bounds for $\varepsilon_{AB} (r_H)$ and that of $\varepsilon_C(r_H)$, we obtain the bound $\|f(\tau)\| \leq \varepsilon_\mr{LR}$ with taking $\varepsilon_\mr{LR}$ by Eq. (\ref{Aeq:Result_epsilon_LR}), thereby completing the proof of Lemma \ref{Alemma:Hamiltonian_restriction}. \hfill $\quad \square$

Let us discuss in what conditions we can extend our results to other cases. 
The change in the dimension and the range of interactions only affects the proper choice the partial system size $L^\prime$, which designates the linear scale of the Hamiltonian restriction. 
Once $L^\prime$ is determined, the remaining protocol is completely same as that of the one-dimensional finite-ranged cases; we compile the dynamics using a quantum system with size $\tilde{L} \geq L^\prime + 2d^\prime +1$ [$d=L^\prime/4+d^\prime$: depth of the variational quantum circuit $V(\theta)$]. 
Therefore, it is sufficient to make $\varepsilon_\mr{LR}$ small enough with a proper size $\tilde{L}$ based on Theorems \ref{Thm:local_compilation_PBC} and \ref{Thm:Local_compilation_generic}.
Depending on what kind of observables is focused on, we have different conditions.
When considering local observables under the approximate circuit $V^{(L)}(\theta_\mr{opt})$, we require $\varepsilon_\mr{LR} \ll 1 $ to keep the local cost functions small according to Eqs. (\ref{Eq:thm_PBC_result_LHST}) or (\ref{Eq:thm_generic_result_LHST}). In this case, to extend our results, it is thus sufficient to choose sufficiently large $L^\prime$ that makes $\varepsilon_\mr{LR} \ll 1$ while keeping $L^\prime / L < 1$ so that the compilation size is smaller than $L$.
On the other hand, when a near-unity average gate fidelity is required for global observables, we demand that $|\Lambda| \varepsilon_\mr{LR} \sim L^D \varepsilon_\mr{LR} \ll 1$ based on Eqs. (\ref{Eq:average_gate_fidelity_pbc}) and (\ref{Eq:average_gate_fidelity_generic}). As a result, the sufficient condition in that case is to achieve $L^D \varepsilon_\mr{LR} \ll 1$ with sufficiently-large $L^\prime$ while keeping $L^\prime / L <1$.
In the following subsections, we derive how $\varepsilon_{\mr{LR}}$ scales with respect to $L^\prime$ in finite-ranged, short-ranged, and long-ranged interacting cases to confirm that our protocol can be applied to these setups.

\subsection{Finite-ranged cases in generic dimension}\label{Asubsec:Extension_Finite_range}

We consider finite-ranged cases in generic dimension. As introduced in Eq. (\ref{Eq:Finite_range}), we here assume
\begin{equation}
    h_X = 0, \quad \text{if $\,^\exists j, j^\prime \in X$ s.t. $\mr{dist}(j,j^\prime) > d_H$},
\end{equation}
where $d_H$ designates the range of interactions. Finite-ranged interacting systems have the LR bound $\mathcal{C}(r,t) = C \exp { - (r-vt)/\xi}$ under a fixed time $t$, with some constants $C$, $v$, and $\xi$, as introduced in Eq. (\ref{Eq:Lieb_Robinson_Bound}) \cite{Lieb1972-uo}. 

Let us evaluate the bound $\varepsilon_\mr{LR}$. We set $L^\prime=2(l_0+d_H+v\tau)$ with a tunable scale $l_0$, and choose the parameter $r_H$ in Eq. (\ref{Aeq:Result_epsilon_LR}) by $r_H=d_H$ ($\leq L^\prime/2$). From the assumption of the range of interactions, $\varepsilon(r_H)$, defined by Eq. (\ref{Aeq:Result_epsilon_rH}), vanishes. This results in the bound,
\begin{equation}\label{Aeq:Bound_LR_finite_range_integral}
    \varepsilon_\mr{LR} = C_1 \int_{l_0+v\tau}^\infty r^{D-1} e^{-(r-v\tau)/\xi} dr,
\end{equation}
reproducing Eq. (\ref{Eq:Def_epsilon_LR2}) in the main text. With some elementary integration using the gamma functions, we arrive at
\begin{equation}\label{Aeq:Bound_LR_finite_range_polynomial}
    \varepsilon_\mr{LR} = C_1 e^{-l_0/\xi}  \sum_{k=0}^{D-1} \frac{(D-1)!}{(D-1-k)!} (l_0+v\tau)^{D-1-k} \xi^k.
\end{equation}
Since the term in the summation is a polynomial of degree $D-1$ in $l_0+v\tau$, there exists a positive constant $C_4$ satisfying
\begin{eqnarray}
    \varepsilon_\mr{LR} &\leq& C_4 (l_0+v\tau)^{D-1} e^{-l_0/\xi} \nonumber \\
    &=& C_4 \exp \left\{ -l_0 / \xi + (D-1) \log (l_0+v\tau) \right\}. \nonumber \\
    && \label{Aeq:Bound_LR_finite_range}
\end{eqnarray}

Since $\varepsilon_\mr{LR}$ exponentially decays in $l_0$ with polynomial corrections,  both $\varepsilon_\mr{LR}$ and $L^D \varepsilon_\mr{LR}$ can be arbitrarily small with sufficiently large $L^\prime$ such that $L^\prime / L < 1$. 
Thus, we can apply the LVQC protocol to finite-ranged cases including high-dimensional systems.

Next, let us discuss how to choose the appropriate compilation size $\tilde{L}$.
When focusing on local observables, we demand $\varepsilon_\mr{LR} \ll 1$, which results in the following choice;
\begin{enumerate}
    \item Choose $l_0$ so that
    \begin{equation}\label{Aeq:Finite_range_term}
        \exp \left\{ -l_0 / \xi + (D-1) \log (l_0+v\tau) \right\}
    \end{equation}
    can be ignored compared to $1$.
    \item Choose the compilation size by $\tilde{L} = 2 \lceil l_0 + d_H + v\tau + d^\prime + 1/2 \rceil $.
\end{enumerate}
To make Eq. (\ref{Aeq:Finite_range_term}) small enough, $l_0$ should be at least larger than $\xi$, which is the localization length of the LR bound. Thus, our protocol typically requires the linear scale $\tilde{L} \gtrsim 2(\xi+d_H+v\tau+d^\prime)$ for evaluating the cost functions.
High-dimensional cases with $D \geq 2$ have logarithmic corrections in its exponent. Although larger linear scale is required compared to one-dimensional cases, still we can expect much decrease in the size.

On the other hand, when considering global observables, we demand $L^D \varepsilon_\mr{LR} \ll 1$. This brings an additional exponent $D \log L$ to Eq. (\ref{Aeq:Finite_range_term}). 
As a result, the typical size for compilation becomes $\tilde{L} \gtrsim 2(\xi+d_H+v\tau+d^\prime + D \xi \log L)$ to ensure high average gate fidelity for larger quantum systems.

\subsection{Short-ranged cases in generic dimension}\label{Asubsec:Extension_Short_range}
Let us discuss short-ranged interacting systems in generic dimensions. In these cases, the range of interactions is infinite but their strength is suppressed exponentially in the distance as
\begin{equation}
    \sum_{X; X \ni j, j^\prime} \| h_X \| \leq h \exp \left( - \mr{dist}(j,j^\prime) / \zeta \right), \quad \text{$\,^\forall j, j^\prime \in \Lambda$},
\end{equation}
with some positive constants $h$ and $\zeta$, for the Hamiltonian $H^{(L)}=\sum_X h_X$. The LR bound is the same as that of finite-ranged cases, $\mathcal{C}(r,t) = C \exp { - (r-vt)/\xi}$ \cite{Robinson1976-wd,Nachtergaele2006-ok,Nachtergaele2006-il}.

Now, we evaluate the bound $\varepsilon_\mr{LR}$ for short-ranged cases. We choose the size $L^\prime$ by $L^\prime = 2(l_0 + r_H + v\tau)$ with two tunable parameters $l_0$ and $r_H$. The first term of $\varepsilon_\mr{LR}$ in Eq. (\ref{Aeq:Result_epsilon_LR}) is the same as that of finite-ranged cases, resulting in the bound in Eq. (\ref{Aeq:Bound_LR_finite_range}). 
The second term $\varepsilon(r_H)$ is then bounded by
\begin{eqnarray}
\varepsilon(r_H) &\leq& C_2 \sum_{i \in \Lambda_{L^\prime,j}} \sum_{i^\prime \in \Lambda; \mr{dist}(i,i^\prime)>r_H} \sum_{X; X \ni i, i^\prime} \|h_X \| \nonumber \\
&\leq& C_2 h \sum_{i \in \Lambda_{L^\prime,j}} \sum_{i^\prime \in \Lambda; \mr{dist}(i,i^\prime)>r_H} \exp \left( - \mr{dist}(i,i^\prime) / \zeta \right). \nonumber \\
&&
\end{eqnarray}
We can again replace the summation over $i$ and $i^\prime$ by the integration over the $D$-dimensional real space like the derivation of Eq. (\ref{Aeq:Integral_LR_bound}) from Eq. (\ref{Aeq:Bound_epsilon_AB}). With the use of a proper positive constant $C_5$, independent of $L$ and $L^\prime$, we arrive at the following bound;
\begin{equation}\label{Aeq:epsilon_rH_short_range}
    \varepsilon(r_H) \leq C_5 (L^\prime)^D (r_H)^{D-1} e^{-r_H/\zeta}.
\end{equation}

Finally, using the relation $L^\prime = 2 (l_0+r_H+v\tau)$, $\varepsilon_\mr{LR}$ satisfies the following inequality;
\begin{eqnarray}
\varepsilon_\mr{LR} &\leq& C_4 (l_0+v\tau)^{D-1} e^{-l_0/\xi}  \nonumber \\
&\quad& + C_6 (l_0+r_H+v\tau)^D r_H^{D-1} e^{-r_H/\zeta},
\end{eqnarray}
where $C_4$ and $C_6$ are some positive constants independent of $L$ and $L^\prime$. 

Similar to finite-ranged cases, both $\varepsilon_\mr{LR}$ and $L^D \varepsilon_\mr{LR}$ can be arbitrarily small with properly increasing $l_0$ and $r_H$ under $L^\prime / L < 1$. 
When we focus on local observables for larger-scale dynamics demanding $\varepsilon_\mr{LR} \ll 1$, we should choose the compilation size $\tilde{L}$ in the following way.
\begin{enumerate}
    \item Choose $l_0$ so that
    \begin{equation}\label{Aeq:Short_range_first}
        \exp \left\{ -l_0 / \xi + (D-1) \log (l_0+v\tau) \right\}
    \end{equation}
    can be ignored compared to $1$.
    \item Choose $r_H$ so that
    \begin{equation}\label{Aeq:Short_range_second}
        \exp \left\{ - \frac{r_H}{\zeta} + D \log (l_0+r_H+v\tau) +(D-1) \log r_H \right\}
    \end{equation}
    can be ignored compared to $1$, under the above choice of $l_0$.
    \item Choose the compilation size by $\tilde{L} = 2 \lceil l_0 + r_H + v\tau + d^\prime + 1/2 \rceil $.
\end{enumerate}

In contrast to finite-ranged cases, the error $\varepsilon_\mr{LR}$ always has logarithmic corrections in its exponent, and has two independent tunable parameters for the scale $\tilde{L}$. 
To make both Eqs. (\ref{Aeq:Short_range_first}) and (\ref{Aeq:Short_range_second}) sufficiently small, the compilation size $\tilde{L}$ should be at least larger than $2(\xi+\zeta+v\tau+d^\prime)$ ($\zeta$: the typical range of interactions), which gives the typical size scale of short-ranged cases.
When the high average gate fidelity is required, we replace the protocol by adding $D \log L$ to the exponents of Eqs. (\ref{Aeq:Short_range_first}) and (\ref{Aeq:Short_range_second}), to achieve $L^D \varepsilon_\mr{LR} \ll 1$. 
Then, the typical compilation size scale becomes $\tilde{L} \gtrsim 2\{ \xi+\zeta+v\tau+d^\prime + D (\xi + \zeta) \log L \}$.

\subsection{Long-ranged cases in generic dimension}\label{Asubsec:Extension_Long_range}
The last case we consider is a long-ranged Hamiltonian in generic dimension. Here, we assume power-law interactions, satisfying
\begin{equation}\label{Aeq:Def_Long_range}
    \sum_{X: X \ni j, r(X) \geq R} \| h_X \| \leq \frac{h}{R^\alpha}, \quad \,^\forall j \in \Lambda,
\end{equation}
for any sufficiently large distance $R > 0$, where  $h$ and $\alpha$ denote some positive constants.
One of the simplest cases is the long-ranged transverse Ising model defined by
\begin{equation}
    H = \sum_{j,j^\prime \in \Lambda, j \neq j^\prime} \frac{Z_j Z_{j^\prime}}{\mr{dist}(j,j^\prime)^{D+\alpha}} + \sum_{j \in \Lambda} X_j,
\end{equation}
on a $D$-dimensional lattice $\Lambda$.
While a series of recent studies have succeeded in extending the LR bound to long-ranged cases in different ways \cite{Hastings2006-ob,Foss-Feig2015-ca,Matsuta2017-mf,Else2020-tr,Kuwahara2020-se,Tran2021-pv}, we hereby focus on one of their results, derived in Ref. \cite{Else2020-tr}.
When the power $\alpha$ is larger than the dimension $D$, there exist positive constants $v$, $C_7$, and $C_8$, such that
\begin{equation}
    \mathcal{C}(r,\tau) \leq C_7 \exp \left( v\tau - r^{1-\sigma} \right) + C_8 \frac{f_\sigma(v\tau)}{r^{\sigma \alpha}},
\end{equation}
for any $\sigma$ satisfying $(D+1)/(\alpha+1)<\sigma<1$. Here, $f_\sigma(x)$ is a monotonically increasing function in $x$ independent of $L$, and can be regarded as a positive constant for fixed $\tau$ and $\sigma$.

We compute the upper bound of $\varepsilon_\mr{LR}$ based on Eq. (\ref{Aeq:Result_epsilon_LR}). The intermediate size $L^\prime$ is again given by $L^\prime = 2(l_0+r_H+v\tau)$ with two tunable parameters $l_0$ and $r_H$.
Substituting the above LR bound into Eq. (\ref{Aeq:Result_epsilon_LR}), the first term of Eq. (\ref{Aeq:Result_epsilon_LR}) is bounded by
\begin{eqnarray}
    \int_{l_0+v\tau}^\infty r^{D-1} \mathcal{C}(r,\tau) dr &\leq& C_7 e^{v\tau} \int_{l_0+v\tau}^\infty r^{D-1} e^{-r^{1-\sigma}} dr \nonumber \\
    &\qquad& + C_8 f_\sigma(v\tau) \int_{l_0+v\tau}^\infty r^{D-1-\sigma \alpha} dr. \nonumber \\
    && \label{Aeq:Integral_bound_Long_range}
\end{eqnarray}
The first integral in the right hand side is computed by the substitution of $s=r^{1-\sigma}-(l_0+v\tau)^{1-\sigma}$, which results in
\begin{eqnarray}
    && [\text{The first term in the r.h.s of Eq. (\ref{Aeq:Integral_bound_Long_range})}] \nonumber \\
    && = \frac{C_7 e^{v\tau-(l_0+v\tau)^{1-\sigma}}}{1-\sigma} \int_0^\infty \{ s+ (l_0+v\tau)^{1-\sigma} \}^{\frac{D}{1-\sigma}-1} e^{-s} ds \nonumber \\
    && \leq \frac{C_7 e^{v\tau-(l_0+v\tau)^{1-\sigma}}}{1-\sigma} \int_0^\infty \{ s+ (l_0+v\tau)^{1-\sigma} \}^{n_{D\sigma}-1} e^{-s} ds, \nonumber \\
    && \label{Aeq:Bound_gamma_integral_Long_range}
\end{eqnarray}
with $n_{D\sigma} = \lceil D/(1-\sigma) \rceil \in \mathbb{N}$. As we derive Eq. (\ref{Aeq:Bound_LR_finite_range}) from Eq. (\ref{Aeq:Bound_LR_finite_range_integral}) using the gamma functions, there exists a positive constant $C_9$, which is dependent only on $D$ and $\sigma$, such that
\begin{eqnarray}
C_1 \times [\text{Eq. (\ref{Aeq:Bound_gamma_integral_Long_range})}] &\leq& C_9 e^{v\tau-(l_0+v\tau)^{1-\sigma}} (l_0+v\tau)^{n_{D\sigma}(1-\sigma)} \nonumber \\
&\leq& C_9 e^{v\tau-(l_0+v\tau)^{1-\sigma}} (l_0+v\tau)^{D+1-\sigma} \nonumber \\
&& \label{Aeq:Integral_bound_Long_range_1st}
\end{eqnarray}
is satisfied. 
On the other hand, considering $D-1-\sigma \alpha < -1$ from $(D+1)/(\alpha+1) < \sigma <1$, the second integral in the right hand side of Eq. (\ref{Aeq:Integral_bound_Long_range}) is easily computed as
\begin{equation}\label{Aeq:Integral_bound_Long_range_2nd}
C_8 f_\sigma(v\tau) \int_{l_0+v\tau}^\infty r^{D-1-\sigma\alpha} dr = \frac{C_8 f_\sigma (v\tau)}{\sigma \alpha - D} (l_0+v\tau)^{D-\sigma \alpha}.
\end{equation}
We define a positive constant $C_{10}$ by $C_{10}=C_8 f_\sigma(v\tau)/\{C_1 (\sigma\alpha-D) \}$, and then Eqs. (\ref{Aeq:Integral_bound_Long_range_1st}) and (\ref{Aeq:Integral_bound_Long_range_2nd}) imply
\begin{eqnarray}
&& [\text{The first term of $\varepsilon_\mr{LR}$ in Eq. (\ref{Aeq:Result_epsilon_LR})}] \leq \nonumber \\
&& C_9 e^{v\tau-(l_0+v\tau)^{1-\sigma}} (l_0+v\tau)^{D+1-\sigma} + C_{10} (l_0+v\tau)^{D-\sigma\alpha}. \nonumber \\
&& \label{Aeq:epsilon_LR_1st_Long_range}
\end{eqnarray}
We note that this bound is independent of $L$, and vanishes with increasing $l_0 \to \infty$.

When the tunable parameter $r_H$ is sufficiently large, the second term $\varepsilon(r_H)$, defined by Eq. (\ref{Aeq:Result_epsilon_rH}), immediately satisfies the following inequality,
\begin{equation}\label{Aeq:epsilon_rH_Long_range}
    \varepsilon (r_H) \leq C_2 |\Lambda_{L^\prime,j}| \cdot \frac{h}{(r_H)^\alpha},
\end{equation}
where we use the assumption of long-range interactions, Eq. (\ref{Aeq:Def_Long_range}). Considering that the volume $|\Lambda_{L^\prime,j}|$ is proportional to $(L^\prime)^D$, there exists a positive constant $C_{11}$ such that $\varepsilon (r_H) \leq C_{11} (l_0+r_H+v\tau)^D \cdot (r_H)^{-\alpha}$. 
From the assumption of $\alpha > D$, this bound vanishes under $r_H \to \infty$ when the other parameter $l_0$ is fixed.

Summarizing the results in Eqs. (\ref{Aeq:epsilon_LR_1st_Long_range}) and (\ref{Aeq:epsilon_rH_Long_range}), we obtain the bound of $\varepsilon_\mr{LR}$ for long-ranged cases in generic dimension as
\begin{eqnarray}
\varepsilon_\mr{LR} &\leq&  C_9 e^{ v\tau - (l_0+v\tau)^{1-\sigma} + (D+1-\sigma) \log (l_0+v\tau)} \nonumber \\
&\qquad& + C_{10} (l_0+v\tau)^{D-\sigma\alpha} + C_{11} \cdot \frac{ (l_0+r_H+v\tau)^D }{(r_H)^\alpha}. \nonumber \\
&& \label{Aeq:epsilon_LR_long_range_result}
\end{eqnarray}
In contrast to finite-ranged and short-ranged cases, the bound $\varepsilon_\mr{LR}$ shows polynomial decay in $\tilde{L}$, which leads to the absence of characteristic length. 
In addition, this also alters applicability of the LVQC protocol depending on which we focus on local or global observables for larger-scale systems.

When we are interested in local observables, $\varepsilon_\mr{LR} \ll 1$ is demanded. Since $\varepsilon_\mr{LR}$ is independent of $L$, we can make $\varepsilon_\mr{LR}$ arbitrarily small by increasing $l_0$ and $r_H$ under the constraint $L^\prime / L < 1$.
We can apply the LVQC protocol as long as the LR bound exists (e.g. $\alpha > D$ is required when we employ the LR bound in Ref. \cite{Else2020-tr}).
The proper compilation size $\tilde{L}$ is organized by the following steps;
\begin{enumerate}
    \item Choose $l_0$ so that both of 
    \begin{equation}
        e^{ v\tau - (l_0+v\tau)^{1-\sigma} + (D+1-\sigma) \log (l_0+v\tau)}
    \end{equation}
    and $(l_0+v\tau)^{D-\sigma \alpha}$ become sufficiently small compared to $1$.
    \item Choose $r_H$ so that $(l_0+r_H+v\tau)^D/(r_H)^\alpha$ can be ignored compared to $1$, under the above choice of $l_0$.
    \item Choose the compilation size by $\tilde{L} = 2 \lceil l_0 + r_H + v\tau + d^\prime + 1/2 \rceil $.
\end{enumerate}
Here, we have options in the parameter $\sigma$ satisfying $(D+1)/(\alpha+1)< \sigma < 1$. Since the constants $C_9$ and $C_{10}$ are divergent for $\sigma$ around its lower and upper bounds [See Eqs. (\ref{Aeq:Bound_gamma_integral_Long_range}) and (\ref{Aeq:Integral_bound_Long_range_2nd})], a possible good choice may be $\sigma = \{ (D+1)/(\alpha+1) +1 \}/2$. 

When we are interested in global observables, we demand $L^D \varepsilon_\mr{LR} \ll 1$. The protocol to choose $L^\prime$ is largely the same as the above one, where each term in $\varepsilon_\mr{LR}$ is replaced by the corresponding term in $L^D \varepsilon_\mr{LR}$.
However, due to the polynomial decay of $\varepsilon_\mr{LR}$ in $l_0$ and $r_H$, we should impose additional conditions on the exponents $\alpha$ and $D$. 
Let us discuss asymptotic behaviour of the compilation size by defining the scaling $l_0 \sim L^\beta$ and $r_H \sim L^\delta$ with $\beta, \delta < 1$. 
Multiplying the right hand side of Eq. (\ref{Aeq:epsilon_LR_long_range_result}) by $L^D$, we have three terms that should decay. 
The first term decays sub-exponentially in $l_0$ but polynomially increases in $L$. It can thus be made arbitrarily small by choosing sufficiently large $l_0$.
With regard to the second term, we demand the convergence of $L^D (l_0+v\tau)^{D - \sigma \alpha} \sim L^{D + \beta (D-\sigma \alpha)}$ (Here we assume $v \tau$ is constant). 
As a result, the inequalities, $\sigma \alpha - D > 0$ and
\begin{equation}
\frac{D}{\sigma \alpha -D} < \beta < 1
\end{equation}
should be satisfied. 
The relation, $\beta < 1$, ensures reduction in the compilation size.
The above inequality implies $\alpha > 2D$ and $\sigma > 2D/ \alpha$ must be satisfied for successful size reduction.
Finally, the third term scales as $L^{D \max (\beta,\delta) - \alpha \delta + D}$. Taking the above constraints on $\beta$ and $\sigma$, the sufficient condition for the vanishing third term is to satisfy
\begin{equation}
    \frac{\sigma D}{\sigma \alpha - D} < \delta < 1.
\end{equation}
To summarize, when demanding the high average gate fidelity, we can apply the LVQC to long-ranged interacting systems with the exponent $\alpha > 2D$, which is stricter than what is required for the existence of the LR bound. 
Then, the compilation size $\tilde{L}$ is at-least proportional to $L^{D/(\sigma \alpha - D)}$ with $2D / \alpha < \sigma < 1$.

Let us finally discuss concrete examples of systems where we can apply LVQC successfully.
With the usage of the LR bound for long-ranged cases derived in Ref. \cite{Else2020-tr}, the constraint for local observables, $\alpha > D$, tells us the availability of the LVQC to various systems, such as 1d systems with dipole-type interactions ($\alpha=2, D=1$) and 1d/2d systems with van der Waals interactions ($\alpha=5,D=1$ or $\alpha=4,D=2$). 
On the other hand, the constraint on global observables, $\alpha > 2D$, implies the applicability to limited cases, such as  1d systems with van der Waals interactions ($\alpha=5,D=1$) within the above examples.
In both cases, the application to long-ranged Hamiltonians of electrons from first-principles (i.e. $\alpha=1-D$ by Coulomb potentials) seems to be difficult with the current knowledge of the LR bound. 
Anyway, we expect applicability of the LVQC to broader systems with the usage of other formulations on the LR bound \cite{Hastings2006-ob,Foss-Feig2015-ca,Matsuta2017-mf,Kuwahara2020-se,Tran2021-pv} or as its further development.

\section{Relation to DQC1-hardness of computing cost functions}\label{Asec:dqc1}

In this section, we discuss how the LVQC protocol is related to computational complexity of QAQC. 
According to Ref. \cite{Khatri2019-gj}, the determination of the cost functions belongs to DQC1-hard problems.
This indicates that efficient QAQC by classical computers is difficult.
On the other hand, our LVQC enables efficient evaluation of the cost functions with a restricted size $\tilde{L}$, and in some cases, we can efficiently complete the protocol by MPS like Sec. \ref{Sec:Numerical}. Here, we resolve this apparent contradiction.

We first introduce the complexity class, DQC1 (deterministic quantum computation with one clean qubit) \cite{Knill1998-xs}. 
Here, we concentrate on a one-dimensional system (extension to higher-dimensional systems is straightforward).
In the DQC1 model, we prepare an $(L+1)$-qubit initial state, composed of one clean qubit and the other qubits lying in a maximally-mixed state, as
\begin{equation}
    \rho = \ket{0} \bra{0} \otimes \left( \frac{\ket{0}\bra{0} + \ket{1} \bra{1}}{2} \right)^{\otimes L}.
\end{equation}
Then, we apply a unitary gate $U$ with the depth up to $\mr{poly}(L)$, and obtain the following probability by measuring the first clean qubit,
\begin{equation}
    p_z = \mr{Tr} [ (\ket{z}\bra{z})_1 U \rho U^\dagger], \quad z = 0,1.
\end{equation}
We refer to the problem of determining the probability $p_z$ with a multiplicative error $\varepsilon < 1$ as the DQC1 models. 
DQC1 models are originally introduced to evaluate the power of nuclear magnetic resonance quantum computes. Famous examples of DQC1-complete problems are estimating spectral density \cite{Knill1998-xs}, trace of unitary matrices \cite{Shepherd2006-ch}, and the Jones polynomials \cite{Shor2007-pc}.
Importantly, Ref. \cite{Fujii2018-yk} proves that, if the probability $p_z$ can be sampled with $\mr{poly}(L)$-time classical algorithms, the polynomial hierarchy will collapse to the second level. 
This implies that efficiently simulating the DQC1 models in classical ways is unlikely.
Recently, Ref. \cite{Khatri2019-gj} has revealed that the determination of the global cost function $C_\mr{HST}$ or the local one $C_\mr{LHST}$ with an error $\varepsilon < 1/ \mr{poly}(L)$ is DQC1-hard for $\mr{poly}(L)$ depth unitaries $U$ and $V$; any DQC1 model can be reduced to the problem of determining the above cost functions.
Based on this fact, quantum compilation with the cost functions $C_\mr{HST}$ or $C_\mr{LHST}$ is also expected to be difficult by classical computation. 

The LVQC seems to give contradictory results by the size reduction. 
Let us consider one-dimensional systems with finite-ranged interactions, and assume that the compilation size $\tilde{L} = 2\lceil l_0 + d_H + v\tau + d^\prime +1/2 \rceil$ satisfies $\tilde{L} \propto \log L$. 
We can classically compute the cost function $C^{(\tilde{L})}(\theta)$ with accuracy $1/\mr{poly}(L)$ by employing matrices whose dimension is $e^{O(\tilde{L})} \sim \mr{poly} (L)$ based on Eqs. (\ref{Eq:Cost_LHST_j}) and (\ref{Eq:Local_cost_generic_case}). It takes at-most $\mr{poly}(L)$ time for its classical evaluation.
Considering that $\varepsilon_\mr{LR}$ is suppressed as $\varepsilon_\mr{LR} < e^{-O(\tilde{L})} = 1/\mr{poly}(L)$, Propositions \ref{Prop:Result1} and \ref{Prop:Result2} (or the proof for Theorem \ref{Thm:Local_compilation_generic}) say
\begin{equation}
    |C_\mr{LHST}(U^{(L)}, V^{(L)}(\theta)) - C^{(\tilde{L})}(\theta)| < \frac{3}{4} \varepsilon_\mr{LR} = 1/ \mr{poly}(L).
\end{equation}
Therefore, we can classically determine the local cost function $C_\mr{LHST} (U^{(L)}, V^{(L)} (\theta))$ with polynomial time in the system size $L$.
Does this imply the collapse of the polynomial hierarchy or the fault of the LVQC formalism?
As the discussion below, the LVQC protocol concludes neither of them.

We resolve the discrepancy depending on the size of the causal cones brought by the LR bound, $v \tau$.
The first case is where the time $\tau$ is constant.
Then, the time evolution operator $U^{(L)}= e^{-iH^{(L)} \tau}$ is not universal under the locality.
The LR bound allows to regard it as a $O(L^0)$-depth circuit in terms of the local observable $C_\mr{LHST}$.
Therefore, while the local cost function  $C_\mr{LHST} (U^{(L)}, V^{(L)} (\theta))$ can be actually obtained by $\mr{poly}(L)$-time classical computation, this case is not problematic. 
The second case is $v \tau \propto L^\kappa$, where we can expect the size reduction if we assume $0 < \kappa < 1$. 
In that case, the compilation size $\tilde{L} = 2\lceil l_0 + d_H + v\tau + d^\prime +1/2 \rceil$ is proportional to $L^\kappa$, and cannot scale as $\log L$. 
Thus, the above discussion predicting the $\mr{poly}(L)$-time classical evaluation is precluded, which results in the consistency of the LVQC with the DQC1-hardness of determining the cost function $C_\mr{LHST}$.
Similarly, the LVQC appears to allow classically-efficient evaluation of the global cost function $C_\mr{HST}$, but there exists no conflict with its DQC1-hardness.

We emphasize some points through this discussion. 
First, in some cases, there remains possibility of the local compilation by classical computers.
For finite-ranged or short-ranged interacting systems under $v \tau = O(L^0)$, the LVQC can be completed with $\mr{poly}(L)$-time classical computation. 
While we employ an approximate classical algorithm relying on MPS in Sec. \ref{Sec:Numerical}, we expect that high-performance classical computers in the future will achieve the compilation for the size $\tilde{L} \sim \log L$ without resorting to any approximation.
On the other hand, we also note that intermediate-scale quantum devices still play a significant role in the local compilation.
While the compilation size $\tilde{L}$ scales as $\log L$ in the above cases under $L \to \infty$, the remaining constant term is not so small for current classical computers.  
For instance, as the numerical simulation in Sec. \ref{Sec:Numerical}, a typical 1d spin chain with finite-range interactions requires $\tilde{L}=20$, resulting in the compilation using $40$-qubit quantum systems.
It will be necessary to prepare hundreds or thousands of qubits for higher-dimensional systems involving finite-, short-, and long-ranged interactions.
Since the DQC1-hardness denies $\mr{poly}(\tilde{L})$-time classical simulation of the local compilation, NISQ devices will be essential to compile larger-scale time evolution operators.

\end{document}